\documentclass{article}

\usepackage{PRIMEarxiv}

\usepackage[utf8]{inputenc}
\usepackage[T1]{fontenc}
\usepackage{hyperref}
\usepackage{url}
\usepackage{booktabs}
\usepackage{amsfonts}
\usepackage{nicefrac}
\usepackage{microtype}
\usepackage{lipsum}
\usepackage{fancyhdr}
\usepackage{graphicx}
\usepackage{subcaption}
\usepackage{amsmath}
\usepackage{tabularx}
\usepackage{xcolor}
\usepackage{multirow}
\usepackage{mathtools}
\usepackage[
  backend=biber,
  style=numeric,
  maxcitenames=2,
  mincitenames=1,
  maxbibnames=10
]{biblatex}
\graphicspath{{media/}}

\title{Revealing the Challenges of Attention-FFN Disaggregation for Modern MoE Models and Hardware Systems}

\author{
  \textbf{Guowei Liu}$^{1}$,~
  \textbf{Hongming Li}$^{1}$,~
  \textbf{Yaning Guo}$^{1}$,~
  \textbf{Yongxi Lyu}$^{1}$,~ \\
  \vspace{-0.1in} \\
  \textbf{Mo Zhou}$^{1}$,~
  \textbf{Yi Liu}\thanks{This work was performed while the author was with Baidu Inc.}~,~
  \textbf{Zhaogeng Li}$^{1}$,~
  \textbf{Yanpeng Wang}$^{1}$~ \\
  \vspace{-0.1in} \\
  {\normalsize $^{1}$Baige AI Team, Baidu Inc.}\\
  \vspace{-0.1in} \\
  \texttt{liuguowei03@baidu.com}
}

\begin{document}
\maketitle

\begin{abstract}
Deploying large-scale MoE models presents challenges in memory capacity and bandwidth for expert activation. While Attention-FFN Disaggregation (AFD) has emerged as a potential architecture to decouple compute and memory resources, its performance boundaries compared to standard large-scale Expert Parallelism (EP) remain underexplored. In this paper, we conduct a systematic analysis of AFD by extending the roofline model to the communication level, correlating interconnect bandwidth, arithmetic intensity, and Hardware FLOPS Utilization (HFU). Our analysis reveals a ``dead zone'' on standard clusters: increasing FFN instance count fails to improve HFU as computational workload is capped by scale-out bandwidth, causing operator active time to shrink relative to the fixed latency budget. We further show that AFD's discrete node-level scaling incurs higher imbalance penalties than EP's continuous batch adjustment. Nevertheless, these limitations diminish under specific conditions: Superpod-class hardware with abundant interconnect bandwidth and models with coarse-grained experts and lower sparsity are more likely to benefit from AFD. These findings position AFD as a promising approach for specific hardware-model combinations rather than a universal solution.
\end{abstract}

\section{Introduction}

The emergence of models with hundreds of billions or even trillions of parameters, such as DeepSeek-V3 (671B parameters) \cite{deepseekai2025deepseekv3technicalreport} and Kimi-K2 (1T parameters) \cite{kimiteam2025kimik2openagentic}, has placed stringent demands on the inference infrastructure of Large Language Models (LLMs). The massive parameter counts and memory requirements (e.g., KV Cache) necessitate parallel inference instances consisting of dozens of GPUs. This vast system scale translates to substantial GPU compute costs, making the optimization of Hardware FLOPS Utilization (HFU) a critical task for improving cost-efficiency. Inference performance optimization is often constrained by Service-Level Objectives (SLO), which become a primary factor limiting model inference concurrency.

In the inference phase, the generation of the first token (i.e., the Prefill stage) and the subsequent tokens (i.e., the Decode stage) exhibit significant differences in FLOPS and memory access requirements. This insight has made Prefill-Decode disaggregation (PD disaggregation) \cite{zhong2024distserve,patel2024splitwise} a standard practice in large-scale deployment, substantially enhancing system throughput. Such a deployment strategy shifts the performance bottleneck of the attention stage in models adopting the MLA architecture, such as DeepSeek-V3, from being memory-bound to being compute-dominated, with this effect being particularly pronounced in the prefill stage. However, due to the auto-regressive nature of the Decode stage, the extremely sparse MoE layers leave the FFN stage constrained by memory bandwidth and communication overhead. This arithmetic intensity insufficiency has prompted the further application of large-scale Expert Parallel (EP) \cite{deepseekai2025deepseekv3technicalreport,largescaleep2025} to Decode instances on top of PD disaggregation. The scheme increases the average tokens per expert during the MoE stage by expanding the DP scale and aggregating batches from multiple DP domains into a single EP batch. While large-scale EP deployment addresses the computational efficiency of the MoE stage, it simultaneously introduces severe DP imbalance and EP imbalance issues.

Attention-FFN Disaggregation (AFD) has emerged as an attractive alternative \cite{zhu2025megascale,step3system}. AFD represents an architectural evolution in current deployment scenarios: through physical decoupling, it enables independent resource scaling and targeted optimization for Attention and FFN roles. However, it is frequently viewed in industry and academia as a panacea for arithmetic intensity problems, often overlooking the complex systemic costs it incurs. Existing research tends to optimistically project throughput gains from AFD while lacking in-depth, scenario-based, and model-parameter-aware quantitative analysis regarding the introduced cross-role communication overhead, pipeline bubbles, and sensitivity to load imbalance. Whether this practice of trading system synergy for operator efficiency truly translates into HFU gains is becoming a neglected yet critical topic in large-scale deployment.

By extending the roofline model to the communication level, this paper provides a systematic analysis of AFD's performance characteristics across different hardware configurations. We find that on standard cluster platforms lacking high-bandwidth, fully-interconnected scale-up networks, AFD systems often degrade from compute-bound states into network bottlenecks. In such states, pursuing operator-level efficiency may not translate into overall HFU gains. We further identify the conditions under which AFD becomes advantageous. The main contributions of this paper are as follows:

\begin{itemize}
    \item \textbf{We propose a budget-based HFU upper-bound analysis method for AFD scenarios.} Through this method, we identify the precarious AFD equilibrium points within specific model and hardware configurations. We demonstrate that communication is often the fundamental factor preventing this equilibrium from being reached on standard clusters, while Superpod architectures can alleviate this constraint.
    \item \textbf{We analyze the sensitivity of deployment strategies to latency jitter in real-world inference workflows.} We demonstrate that, under the same levels of DP and EP imbalance, AFD---which introduces discrete scaling by distributing computation across different nodes---exhibits lower tolerance for load imbalance compared to large-scale EP.
    \item \textbf{We synthesize the conditions favorable for AFD deployment.} Based on our analysis, we identify that Superpod-class hardware systems and models with coarse-grained expert configurations are more likely to benefit from AFD, providing guidance for future infrastructure and model co-design.
\end{itemize}

\section{Preliminary}

\subsection{Notations}

\begin{table}[htbp]
\centering
\caption{Summary of notations and their explanations used throughout this paper.}
\label{tab:notations}
\resizebox{\textwidth}{!}{
\begin{tabular}{@{}llll@{}}
\toprule
\textbf{Symbol} & \textbf{Explanation} & \textbf{Symbol} & \textbf{Explanation} \\ \midrule
$B_A$ & Batch size on attention side & $t_c$ & Communication latency \\
$B_F$ & Batch size on FFN side & $t_f$ & FFN computation latency \\
$B_{\text{rank}}$ & Tokens assigned to a rank on FFN side & $t_g$ & Total gap latency: batch preparation and dense layers \\
$B_{\text{ScaleOut}}$ & Scale-out token throughput in fixed time & $t_{dispatch}$ & Dispatch communication latency \\
$B_{\text{ScaleUp}}$ & Scale-up token throughput in fixed time & $t_{combine}$ & Combine communication latency \\
$\mathcal{B}_{\text{mem}}$ & GPU memory bandwidth & $g$ & Number of GPUs per node \\
$\mathcal{B}_{\text{ScaleOut}}$ & Scale-out network bandwidth & $\sigma$ & Stage throughput degradation due to imbalance \\
$\mathcal{B}_{\text{ScaleUp}}$ & Scale-up network bandwidth & $\alpha$ & Overall throughput degradation due to imbalance \\
$H$ & Model hidden size & $\lambda_{\text{EP}}$ & Ratio of $t_a$ to $t_f$ under large-scale EP deployment \\
$L_{accept}$ & Run batch acceptance length & $\lambda_{\text{AFD}}$ & Ratio of $N_A$ to $N_F$ under AFD \\
$N_A$ & Number of A-role nodes & $I$ & Arithmetic intensity \\
$N_F$ & Number of F-role nodes & $M$ & MoE intermediate size \\
$N_{layers}$ & Number of layers (BO-ready) & $S_t$ & Temporal sparsity \\
$N_{experts}$ & Number of MoE experts & $\text{OFU}$ & Operator FLOPS Utilization (during active period) \\
$N_{BO}$ & BO cardinality (at least 3) & $\text{HFU}$ & Hardware FLOPS Utilization \\
$T$ & Target run batch latency & $\text{FLOPS}$ & Floating Point Operations Per Second \\
$t_B$ & Latency budget per micro-batch per stage & $\text{FLOPs}$ & Number of floating point operations \\
$t_\text{G}$ & Latency for grouped GEMM on FFN side & $\text{Mem}$ & Memory access volume required for computation \\
$t_a$ & Attention computation latency & & \\ \bottomrule
\end{tabular}
}
\end{table}

Overall, we use the uppercase letter $B$ to denote the concept of ``batch'', the calligraphic symbol $\mathcal{B}$ for bandwidth, the uppercase $N$ for integers, the lowercase $t$ for latency, and Greek letters such as $\sigma$ for ratios. Some concepts are represented in their textual form, such as $\text{OFU}$. The notations are summarized in Table~\ref{tab:notations}.

\subsection{AFD and Budget under 3BO}

\begin{figure}[t!]
    \centering
    \begin{subfigure}{0.54\textwidth}
        \centering
        \includegraphics[width=1\linewidth]{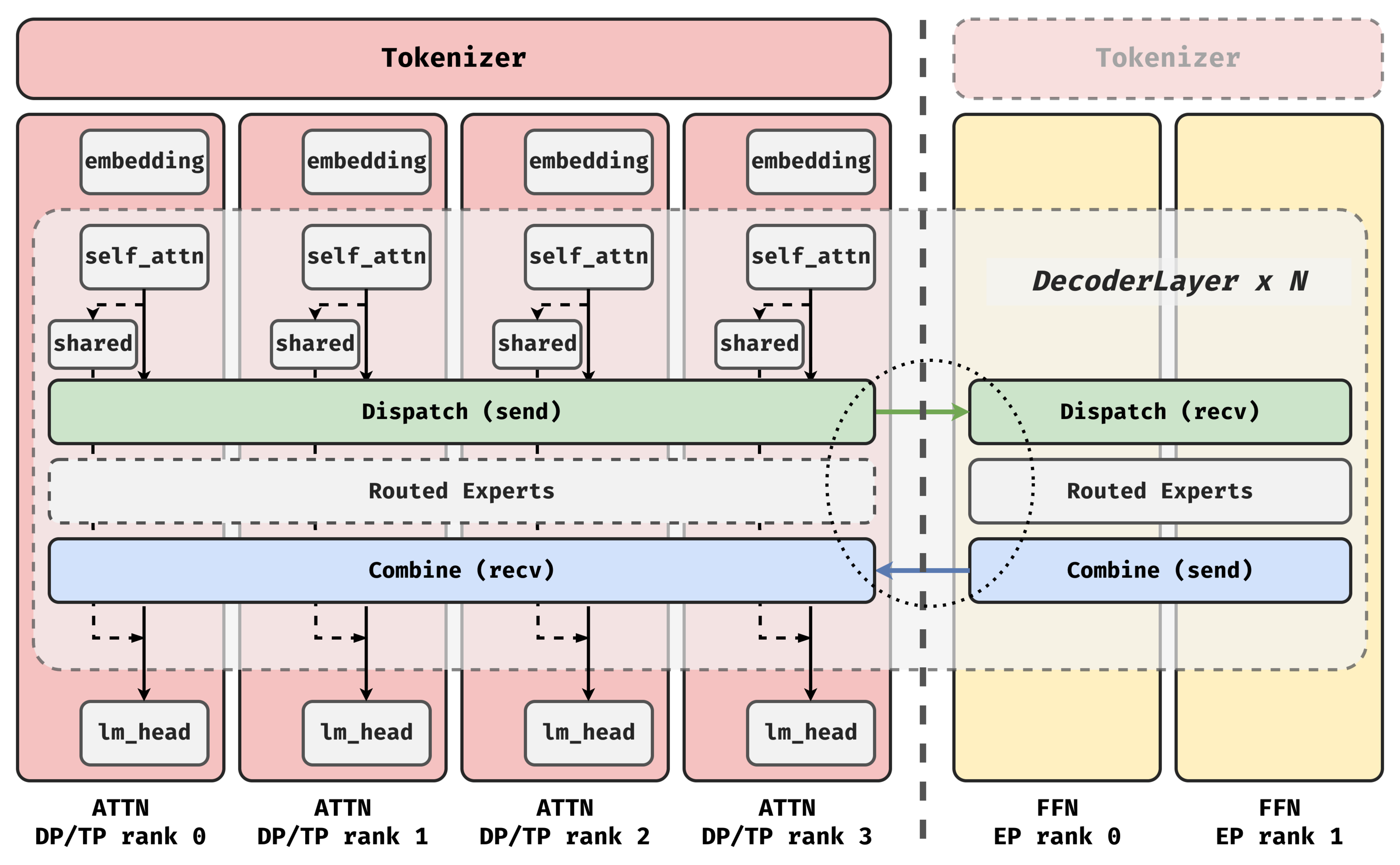}
        \caption{AFD architecture. The attention instances (left) host the majority of the model network, while the FFN instances (right) host the routed experts for all MoE layers. The two instance types are connected via unidirectional collective communication.}
        \label{fig:afd-arch}
    \end{subfigure}
    \hfill
    \begin{subfigure}{0.42\textwidth}
        \centering
        \includegraphics[width=1\linewidth]{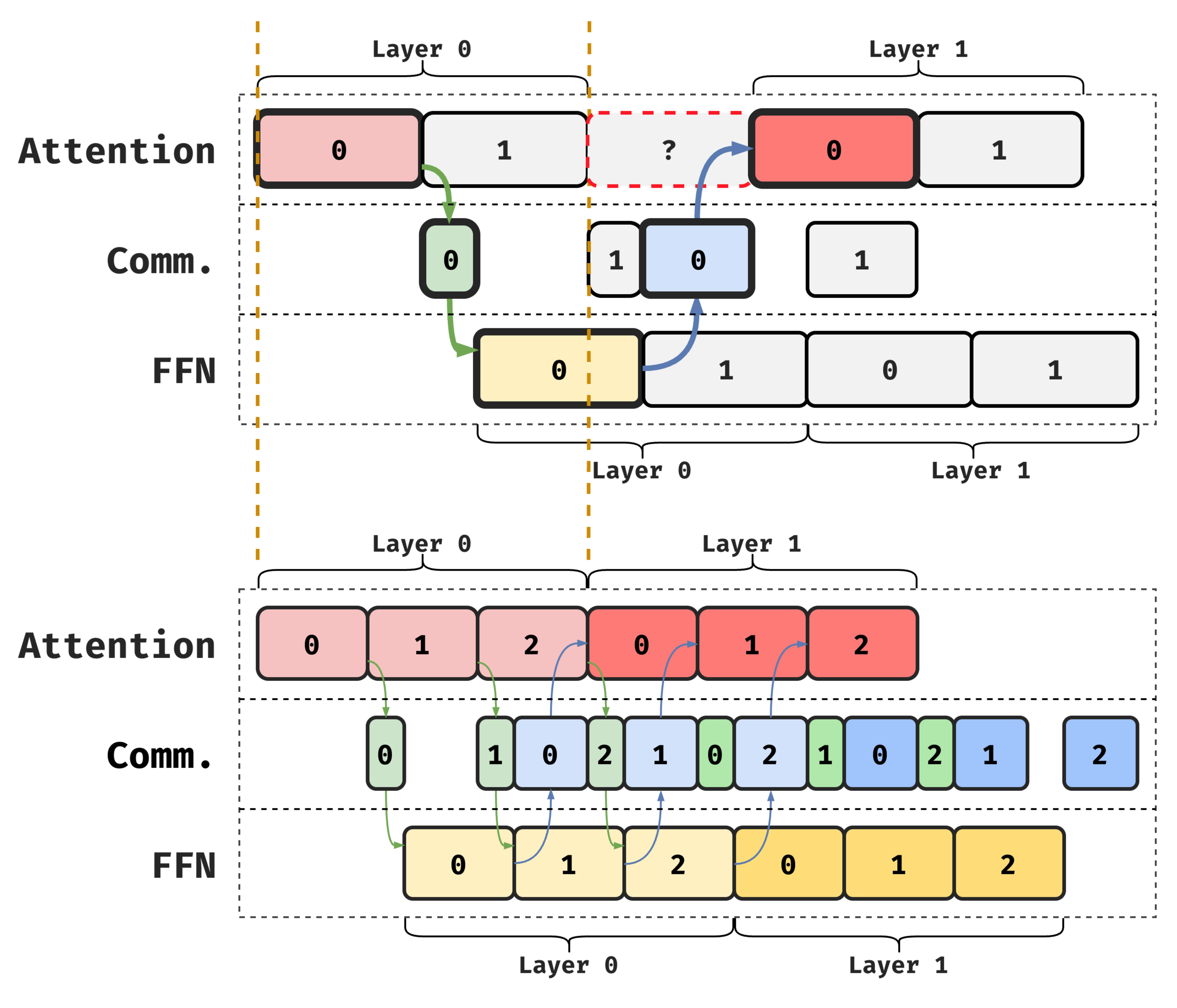}
        \caption{Under similar per-layer latency budgets, 2BO (top) introduces GPU bubbles, while 3BO (bottom) has the potential to maintain a bubble-free pipeline.}
        \label{fig:2bo-bubble-and-3bo}
    \end{subfigure}

    \caption{Illustration of AFD architecture (left) and micro-batch overlap strategies (right). The AFD architecture physically separates attention and FFN computations, while the overlap strategy determines how micro-batches are pipelined to hide communication latency.}
\end{figure}

Figure~\ref{fig:afd-arch} illustrates a typical AFD architecture, where the attention instances integrate the majority of the model network from original Decode instances, while routed experts across all MoE layers are extracted to form new FFN instances with an independent scale. Since shared experts generally operate at high arithmetic intensity, the actual deployment scheme can retain them on attention instances or transfer them to FFN instances as needed. Based on this disaggregation concept, AFD differs from existing large-scale EP deployments in several key aspects.

The first distinction concerns \textbf{communication architectures}. Traditional EP communication, such as DeepEP \cite{deepep2025}, adopts an $\textit{all-to-all}$ pattern where all ranks participate symmetrically in token sending and receiving. In AFD, this pattern transforms into a unidirectional $M2N$ communication, specifically a directed collective communication within bipartite graph semantics. For instance, in the dispatch phase, tokens are transmitted from attention instances to FFN instances based on gating results. Although both the source and destination sides involve multiple GPU ranks, the data flow exhibits unidirectional (attention to FFN) and asymmetric (distinct behaviors for different roles) characteristics.

Another difference is the requirement of at least \textbf{Three-batch overlap (3BO)} in the design of forward propagation under AFD. In current best practices for large-scale EP deployment, No Batch Overlap (NBO), Single-Batch Overlap (SBO), or Two-Batch Overlap (2BO) are selected within Decode instances based on hardware capabilities and batch size. Their main differences are shown in Table~\ref{tab:bo_methods}.

\begin{table}[htbp]
\centering
\footnotesize
\caption{Comparison of Batch Overlap (BO) Methods: Execution Flow, Scenario, and Practices. Micro-batch IDs are denoted in square brackets.}
\label{tab:bo_methods}
\begin{tabularx}{\textwidth}{@{}l c X X c c p{3.5cm}@{}}
\toprule
\multirow{2}{*}{\textbf{Method}} & \multirow{2}{*}{\textbf{Stage}} & \multicolumn{2}{c}{\textbf{Overlap Execution}} & \multicolumn{2}{c}{\textbf{Scenario}} & \multirow{2}{2.5cm}{\textbf{Example Practice}} \\ \cmidrule(lr){3-4} \cmidrule(lr){5-6}
 & & \textbf{Compute Stream} & \textbf{Comm. Stream} & \textbf{FLOPS} & \textbf{Shared Expert(s)} & \\ \midrule
NBO & \#1 & Standard MoE & N/A & Low & w/ or w/o & N/A \\ \addlinespace
SBO & \#1 & Shared Expert & EP Dispatch & Low & w/ & H20 + DeepSeek-V3 \cite{sbopractise2025} \\ \addlinespace
\multirow{2}{*}{2BO} & \#1 & Shared[0] + QKV[1] & EP Dispatch[0] & \multirow{2}{*}{High} & \multirow{2}{*}{w/} & \multirow{2}{\linewidth}{H800 + DeepSeek-V3 \cite{largescaleep2025}} \\
 & \#2 & Attn[1] & EP Combine[0] & & & \\ \addlinespace
3BO & \#1 & Attn $\times$ FFN & All-to-All (EP) & High & w/ or w/o & H800 + Step3 \cite{step3system} \\ \bottomrule
\end{tabularx}
\end{table}

We denote the attention computation, FFN computation, and communication latencies of a micro-batch as $t_a$, $t_f$, and $t_c$, respectively, and the overall single-stage latency budget as $t_B$. Here, $t_c = t_{dispatch} + t_{combine}$ represents the total latency for dispatch and combine communication of a micro-batch. In AFD, \textbf{NBO/SBO is the least preferred option}. AFD inherently possesses separation characteristics, and serial execution would lead to severe device idling, necessitating micro-batch overlap to hide communication overhead. Furthermore, \textbf{2BO inevitably incurs GPU bubbles}. As shown in Figure~\ref{fig:2bo-bubble-and-3bo}, under the same layer count and SLO constraints, the latency budgets for the attention side and FFN side are roughly equivalent layer-wise. Under such an assumption, after a micro-batch is output from attention, the total elapsed time $t_{dispatch} + t_f + t_{combine} > t_a$ inevitably introduces synchronization bubbles in the attention timeline. Consequently, 3BO becomes the minimum requirement to maintain efficient pipeline operation.

It is worth noting that 3BO in AFD, relative to 2BO, is akin to 2BO relative to NBO/SBO in large-scale EP deployment---a promising but not guaranteed bubble-free improvement. In AFD, 3BO provides a higher upper bound while drastically lowering the tolerance for $t_B$ fluctuations. At a lower $t_B$ baseline, once network jitter causes $t_c$ to exceed the budget, or load imbalance causes $t_a$ or $t_f$ to surge, bubbles will rapidly propagate bidirectionally between the two roles, causing more severe performance collapse than in traditional large-scale EP. We detail the analysis on load imbalance in \S\ref{sec:imbalance_penalty}.

In AFD, two roles are responsible for different computational tasks, resulting in two separated operator streams. To satisfy the SLO requirement, letting $T$ denote the run batch latency, we have:

\begin{equation}
    \label{eq:t_b_calculation}
    T = \text{SLO} \times L_{accept} = t_g + N_{layers} \times N_{BO} \times t_{B}
\end{equation}

where $\text{SLO}$ represents the \textit{time-per-output-token} (TPOT) related to the deployment scenario, $t_g$ is the sum of latencies for the inter-batch gap and forward latencies of non-3BO layers, $N_{layers}$ is the number of hidden layers forwarded in 3BO mode, and $N_{BO} = 3$ is the batch overlap cardinality. Thanks to Multi-Token Prediction (MTP) \cite{chen2023accelerating,leviathan2023fast}, we can relax $T$ to $\text{SLO} \times L_{accept}$ instead of $\text{SLO}$, where $L_{accept}$ represents the average acceptance length of a single request in a batch generation. This value mainly depends on whether MTP is enabled, the number of draft tokens, and the statistical average per-token acceptance rate. If the model has no available MTP draft model, then $L_{accept} = 1$. The above formula allows for estimating $t_B$ under scenario and model configurations, thereby assisting in the assessment of attention batch size, FFN batch size, and communication latency.

Equation~\ref{eq:t_b_calculation} presents the upper bound for $t_a$, $t_f$, and $t_c$ without violating the SLO constraint, i.e.,

\begin{equation}
    \max(t_a, t_f, t_c) \le t_B
\end{equation}

However, this does not imply that the AFD system is in an optimal HFU state. To ensure full utilization of the fixed $t_B$ allocated, we aim to optimize such that:

\begin{align}
    2 \times t_a & \ge t_f + t_c \label{eq:t_a_no_bubble} \\
    2 \times t_f & \ge t_a + t_c \label{eq:t_f_no_bubble} \\
    t_a, t_f & \xrightarrow[]{opt.} t_B \label{eq:t_a_t_f_to_t_b}
\end{align}

Here, Equations~\ref{eq:t_a_no_bubble} and~\ref{eq:t_f_no_bubble} represent constraints desired to eliminate GPU bubbles in attention or FFN to enhance utilization, while Equation~\ref{eq:t_a_t_f_to_t_b} expresses the expectation that through optimization, the execution of both attention and FFN layers approaches $t_B$ (and with a high HFU). Ideally, when $t_B = t_a = t_f \ge t_c$, optimal AFD operator occupancy is achieved.

Regarding optimization methods, attention and FFN differ:

\begin{itemize}
    \item For attention, since HFU is nearly in a \textit{compute-bound} state, $t_a \rightarrow t_B$ can be approximated by fine-tuning the batch size of attention ranks.
    \item For FFN, it is necessary to jointly adjust the attention batch, including the parallelism scale and batch size of attention instances, and the scale of FFN instances to pursue $t_f \rightarrow t_B$ in a multi-variable manner.
\end{itemize}

The differences in optimization methods and their complexity dictate that AFD optimization will primarily focus on the FFN side. We desire FFN operators to possess high HFU while maintaining low temporal sparsity (i.e., the ratio of operator-active period to the fixed allocated period).

\subsection{Performance Metrics}
In most scenarios, we can evaluate the system's utilization of computing resources by directly measuring HFU. In this paper, we further introduce the concepts of Operator FLOPS Utilization (OFU)\footnote{We use ``operator'' rather than ``kernel'' to maintain hardware-agnostic terminology throughout this paper.} and temporal sparsity (denoted as $S_t$). We define OFU as the FLOPS utilization of an operator during its active period, equal to the FLOPs achieved divided by the actual execution latency. For scenarios like AFD where each overlap stage is assigned a fixed duration $t_B$, we use $S_t$, whose value equals the operator execution latency divided by $t_B$, as a metric of operator occupancy during $t_B$.

We also need to define arithmetic intensity $I$, the ratio of computation to the memory access required. For the MoE stage, the forward propagation of the MLP network mainly consists of two grouped GEMM and an activation (such as SwiGLU\cite{shazeer2020gluvariantsimprovetransformer}) between them. Since the fused activation operator occupies negligible latency ($10 \mu s$ magnitude) in a typical operator sequence, we represent the computation on EP ranks using grouped GEMM operators only. Assuming all grouped GEMM operations take FP8 format as inputs, for matrices $A \in \mathbb{R}^{\text{M} \times \text{K}}$, $B \in \mathbb{R}^{\text{K} \times \text{N}}$, and $C \in \mathbb{R}^{\text{M} \times \text{N}}$, the computational demand for matrix multiplication $C = AB$ is $\text{FLOPs} = 2 \times \text{M} \times \text{N} \times \text{K}$, while the total memory access is $\text{Mem} = \text{N} \times \text{K}$, neglecting the small proportion of the activation tensors (i.e., the input and output tensors). The arithmetic intensity of a single expert is thus $I = 2M = 2B$, where $B$ is the average input tokens per expert. The arithmetic intensity of a single expert can be naturally generalized to two grouped GEMM operations.

Let $t_{\text{G}}$ denote the total latency to execute two grouped GEMM operations. We define $S_t$, OFU, and HFU by the following formulas:

\begin{align}
    S_t & = \frac{t_\text{G}}{t_B} \label{eq:s_t}\\
    \text{OFU} & = \frac{\text{FLOPs}}{t_{\text{G}}} \label{eq:OFU}\\
    \text{HFU} & = \frac{\text{FLOPs}}{t_{\text{B}}} = \text{OFU} \times S_{t} \label{eq:HFU}
\end{align}

Note the subtle difference in denominators between $\text{OFU}$ and $\text{HFU}$. By increasing the average input tokens per expert for the grouped GEMM operator, i.e., $B$, the direct result is a higher $\text{OFU}$. We also need to consider temporal sparsity $S_t$ to obtain the $\text{HFU}$ of the entire MoE stage.

\subsection{Trends in Modern MoE Architectures}
\label{sec:trends_for_modern_moe}

Models with small parameter counts mostly adopt a single-expert form. Scaling Laws \cite{kaplan2020scaling,clark2022unified,krajewski2024scaling,tian2025towards} indicate that well-trained larger-scale models can further improve performance. Guided by this insight, more models are increasing their parameter counts. To balance convergence speed during training, MoE architectures reduce the FLOPs requirement for training by activating only a small fraction of experts for a single token. On the other hand, as current model sizes have grown to hundreds of billions or even trillions of parameters, the expert granularity (i.e., the ratio of model hidden size $H$ to MoE intermediate size $M$) is becoming finer, and expert sparsity (i.e., the ratio of routed expert count to $\text{TopK}$) is becoming larger \cite{guo2025sonicmoe}. These trends make current model deployment, compared to traditional TP deployment of small models, more inclined towards distributed and large-scale EP to meet the comprehensive demands for MoE residency and inference latency.

The aforementioned two trends in model design bring the following impacts:

\begin{itemize}
    \item \textbf{Reduced computational demand per token ($\text{FLOPs/tok}$) via greater expert sparsity}: Under the same total parameter count, greater expert sparsity determines the activation size in the MoE layer inversely. This conclusion also applies to the inference stage.
    \item \textbf{Reduced computational demand per expert ($\text{tok/expert}$) via finer expert granularity}: Finer expert granularity essentially partitions the computational workload into more discrete units. While the total activation size remains constant, the simultaneous increase in TopK and the total number of experts causes the incoming tokens to be distributed across a wider array of experts. This ``fan-out'' effect significantly dilutes the per-expert token counts.
\end{itemize}
\section{System Analysis}
\label{sec:system-analysis}

\subsection{Arithmetic Intensity in MoE Inference}
\label{sec:arithmetic_intensity}

\begin{figure}[t!]
    \centering
    \includegraphics[width=0.75\linewidth]{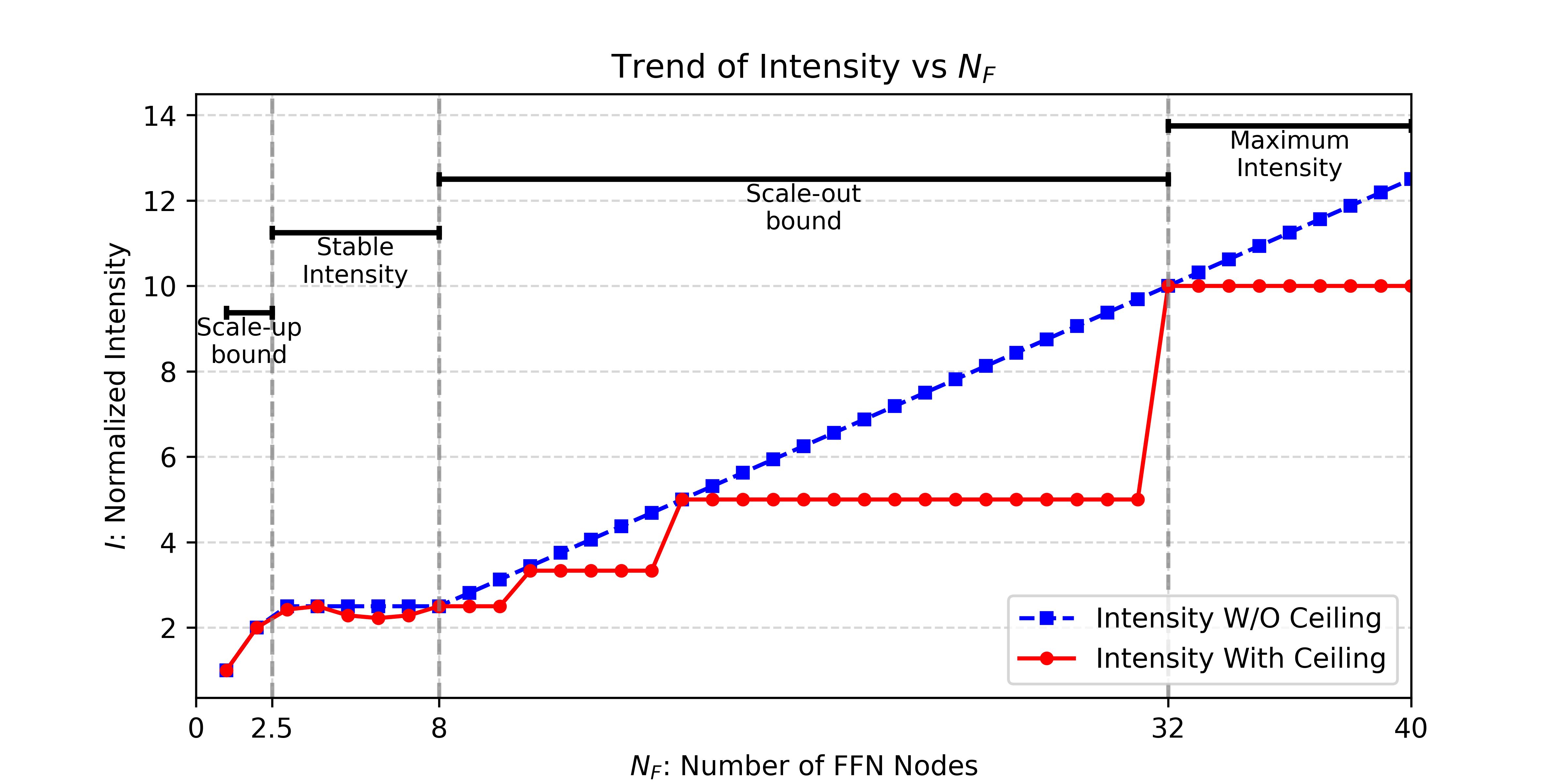}
    \caption{Normalized arithmetic intensity as a function of FFN node count ($N_F$) for DeepSeek-V3 on the H800 platform. The blue curve shows the theoretical upper bound, while the red one indicate actual values accounting for expert count discretization. The four distinct regions correspond to different bandwidth bottleneck regimes.}
    \label{fig:intensity-trend}
\end{figure}

In this section, we examine the impact of model configuration, scale-out network bandwidth (e.g., InfiniBand, RoCE), and scale-up network bandwidth (e.g., NVLink) for AFD bottleneck analysis. To facilitate the discussion, we first limit the scope to the following scenarios and will expand on them later:

\begin{itemize}
    \item The hardware platform is a non-Superpod platform similar to NVIDIA H800.
    \item Attention and FFN instances are deployed at node granularity, i.e., the scale of each instance is a multiple of $g$, the number of GPUs per node.
\end{itemize}

Generally speaking, scale-up network bandwidth is several times the amortized scale-out bandwidth per GPU. For example, NVIDIA H800 offers 160\,GB/s\footnote{200\,GB/s peak per specification; $\le$160\,GB/s sustained in practice; similar assumptions apply throughout.} unidirectional bandwidth ($3.2\times$ that of 50\,GbE/GPU), and NVIDIA H20 offers 360\,GB/s unidirectional bandwidth ($7.2\times$ that of 50\,GbE). Such scale-up networks accelerate data exchange between GPUs within a node, increasing the upper limit of data transmission for AFD inference. In AFD, the design of the communication library generally follows two mainstream implementation approaches:

\begin{itemize}
\item \textbf{Point-to-point direct transfer}: Tokens and their associated metadata are transmitted in a point-to-point manner and written directly into pre-allocated buffers on the destination side, with one buffer reserved for each source rank. Under the AFD setting, this approach cannot effectively leverage high-bandwidth scale-up interconnects.
\item \textbf{Two-stage forwarding}: Taking the \textit{dispatch} phase as an example, in the first stage, tokens and metadata are transmitted over the scale-out network, with each FFN node serving as the transmission target. In the second stage, the data are further forwarded via the scale-up network to the target ranks within the destination node. The \textit{combine} phase follows the reverse logical flow of \textit{dispatch}.
\end{itemize}

Considering the two available communication modes jointly, we measure the overall maximum input token count for a \textbf{single rank} achievable via the interconnect network within $t_B$ by:

\begin{equation}
    B_{\text{rank}} = \min\left(B_{\text{ScaleOut}} \times \max \left(1, \frac{\text{TopK}}{N_F}\right), B_{\text{ScaleUp}}\right)
    \label{eq:b_rank}
\end{equation}

where $B_{\text{ScaleOut}}$ and $B_{\text{ScaleUp}}$ represent the number of tokens that can be transmitted via the scale-out network and scale-up network within $t_B$, respectively, and $N_F$ represents the number of FFN nodes used in the deployment. The term $\max (1, \frac{\text{TopK}}{N_F})$ in the formula represents the pipeline overlap capability of scale-out and scale-up traffic.

Taking DeepSeek-V3 as an example, two 8-GPU H800 nodes are sufficient to support the deployment of a minimal FFN instance. The model's $\text{TopK}$ parameter is 8, which means that an attention output token, in a statistical average sense, may eventually reach $\text{TopK} / N_F = 8 / 2 = 4$ expert input buffers. On the H800 platform equipped with 400\,GbE per GPU, the bandwidth ratio of the scale-up network to the scale-out network is $160 / 50 = 3.2$. Since $\text{TopK} / N_F = 4 > 3.2$, the system is scale-up bound in this configuration, and $B_{\text{rank}} = B_{\text{ScaleUp}} = 3.2 \times B_{\text{ScaleOut}}$.

Let $N_{experts}$ denote the total number of routed experts. The arithmetic intensity $I$ can be expressed as twice the average number of tokens assigned to each expert:

\begin{equation}
    I = \frac{2B_{\text{rank}}}{\lceil N_{experts} / (N_F \times g) \rceil}
    \label{eq:intensity_of_experts}
\end{equation}

\begin{figure}[t!]
    \centering
    \begin{subfigure}{0.48\textwidth}
        \centering
        \includegraphics[width=1.0\linewidth]{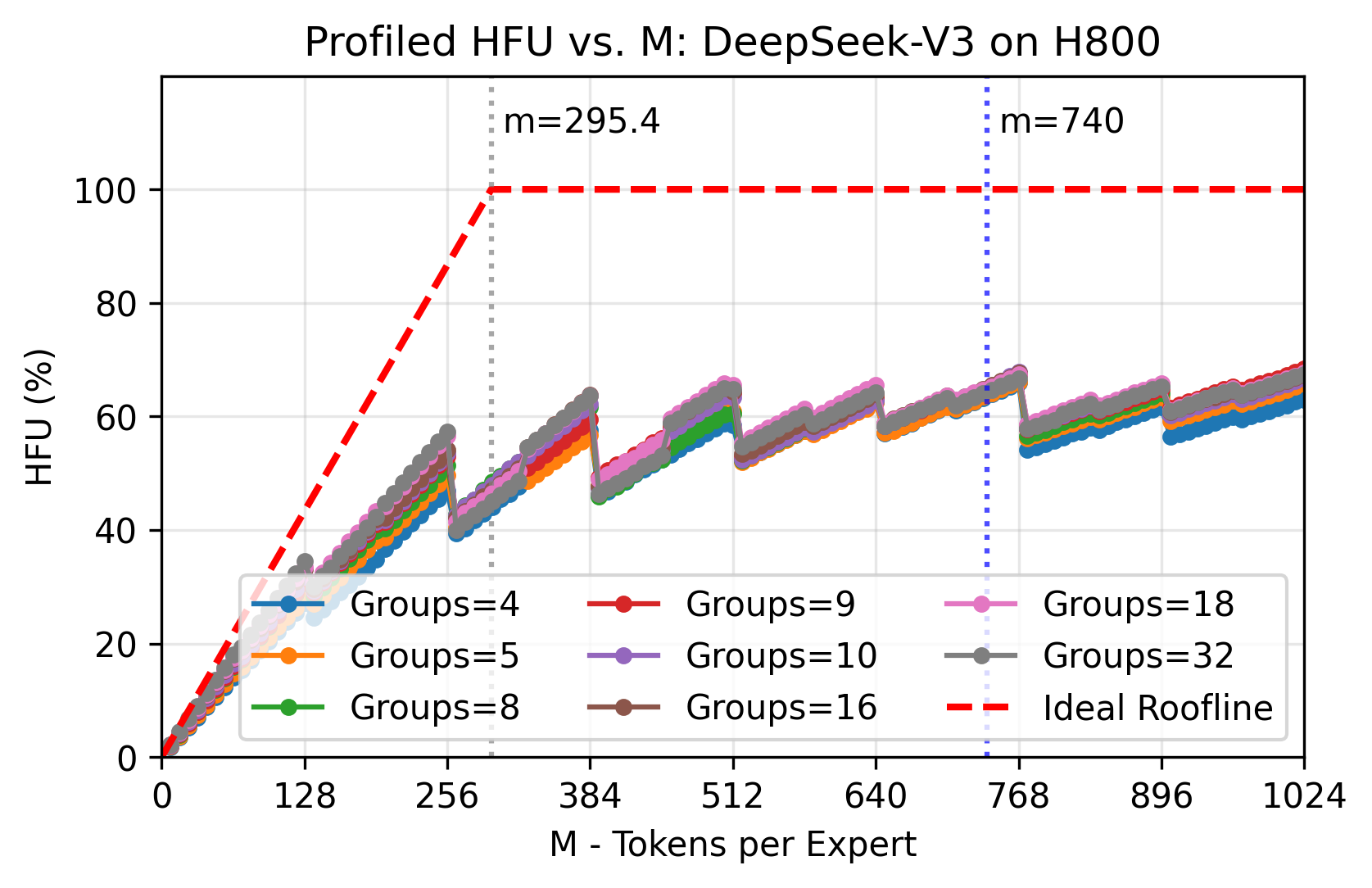}
        \caption{On H800 (balanced token distribution).}
        \label{fig:roofline-h800-balanced}
    \end{subfigure}
    \hfill
    \begin{subfigure}{0.48\textwidth}
        \centering
        \includegraphics[width=1.0\linewidth]{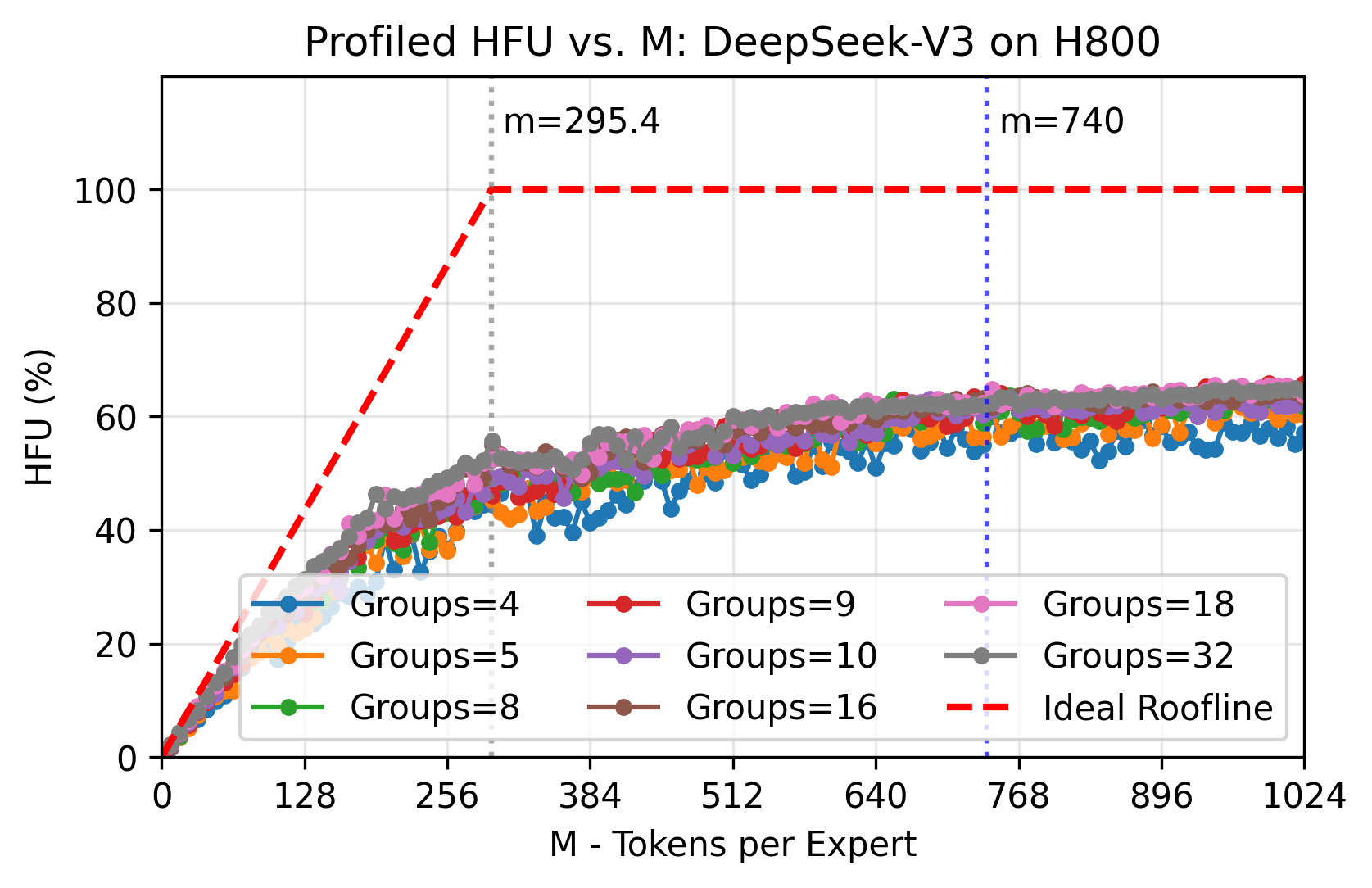}
        \caption{On H800 (imbalanced token distribution).}
        \label{fig:roofline-h800-imbalanced}
    \end{subfigure}

    \vspace{2mm}

    \begin{subfigure}{0.48\textwidth}
        \centering
        \includegraphics[width=1.0\linewidth]{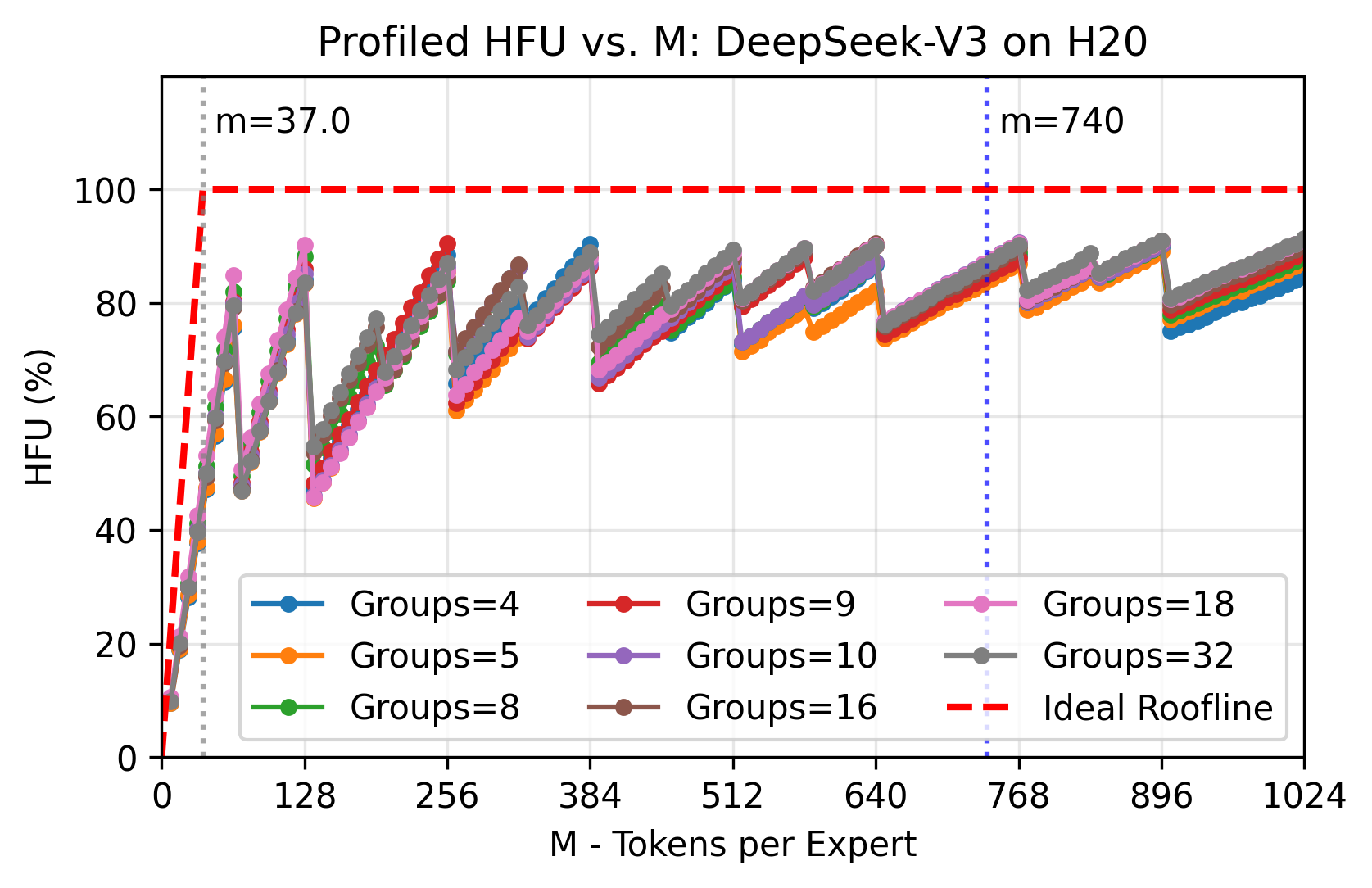}
        \caption{On H20 (balanced token distribution).}
        \label{fig:roofline-h20-balanced}
    \end{subfigure}
    \hfill
    \begin{subfigure}{0.48\textwidth}
        \centering
        \includegraphics[width=1.0\linewidth]{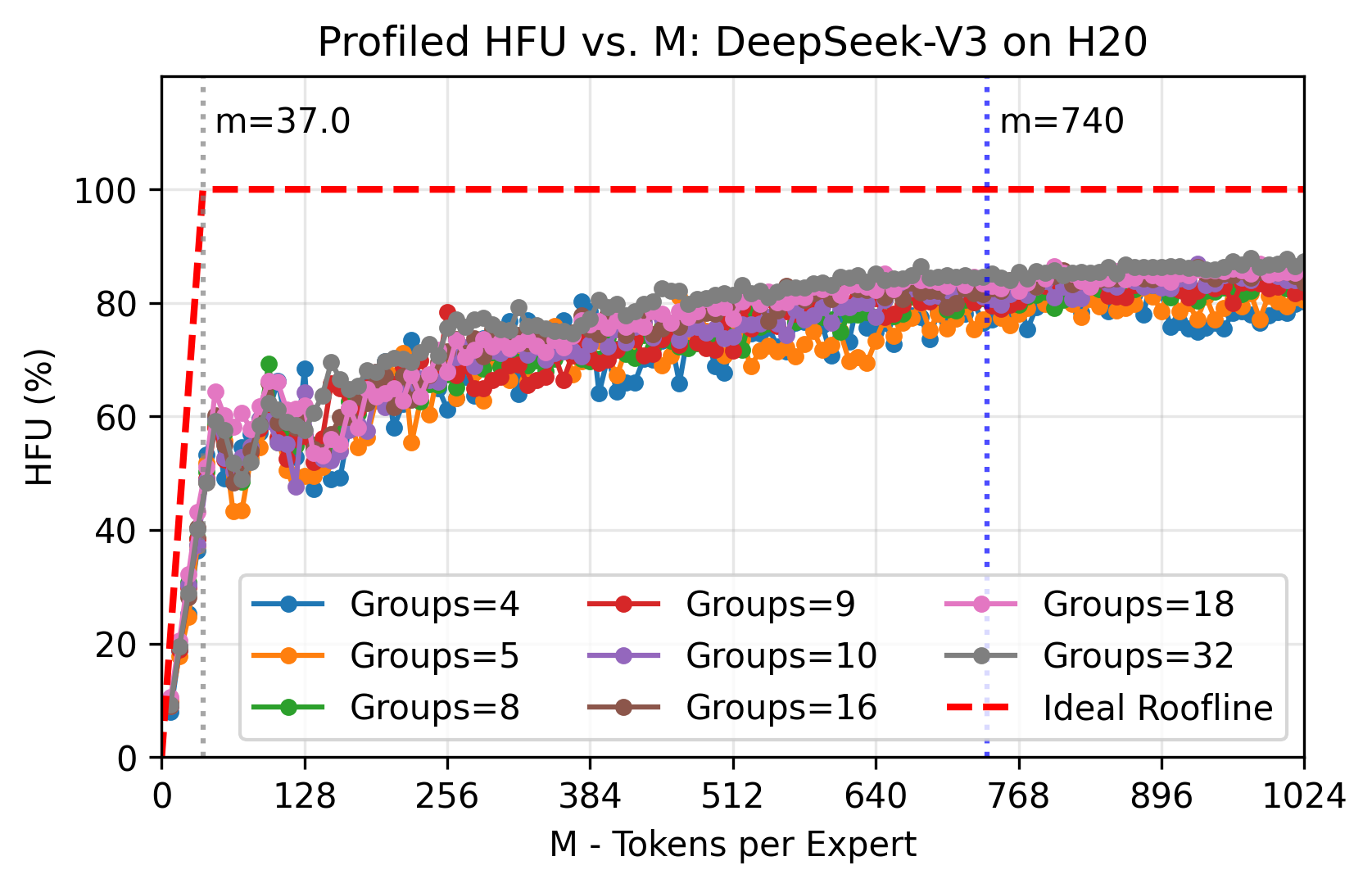}
        \caption{On H20 (imbalanced token distribution).}
        \label{fig:roofline-h20-imbalanced}
    \end{subfigure}

    \caption{Grouped GEMM unit tests and theoretical roofline vs.\ M (average tokens per expert) on H20 and H200 platforms. The left column assumes balanced token distribution across experts, while the right column reflects realistic imbalanced scenarios where some experts receive disproportionately more tokens.}
    \label{fig:roofline}
\end{figure}

We used the DeepSeek-V3 model configuration and H800 platform hardware parameters to plot the trend of normalized arithmetic intensity changing with $N_F$, as shown in Figure~\ref{fig:intensity-trend}. The blue curve represents the upper bound of calculated $I$ without considering the discretization of the local expert count. Based on the value of $N_F$, the relationships defined in Equations~\ref{eq:b_rank} and~\ref{eq:intensity_of_experts} can be partitioned into four distinct operational regimes:

\begin{itemize}
    \item \textbf{Scale-up bound region}: When $N_F$ is small enough such that $\text{TopK} / N_F > B_{\text{ScaleUp}} / B_{\text{ScaleOut}}$, the number of inbound tokens is mainly constrained by the scale-up network bandwidth. Increasing $N_F$ cannot increase $B_{\text{rank}}$. However, since the number of local experts decreases as $N_F$ increases, $I$ improves with $N_F$.
    \item \textbf{Stable intensity region}: When $1 \le \text{TopK} / N_F \le B_{\text{ScaleUp}} / B_{\text{ScaleOut}}$, the number of inbound tokens is constrained by the scale-out bandwidth. Since the number of local experts decreases proportionally with $N_F$, the expected $I$ remains unchanged.
    \item \textbf{Scale-out bound region}: When $N_F \ge \text{TopK}$, the communication stage of AFD cannot benefit from the scale-up network. At this point, the FFN batch size (as well as $I$) is completely constrained by the scale-out bandwidth. Increasing $N_F$ can continue to optimize $I$ by reducing the number of experts that need to be deployed locally.
    \item \textbf{Maximum intensity region}: Constrained by the scale-out network, when $N_F$ continues to increase until the minimum number of local experts drops to 1, increasing $N_F$ will no longer yield larger $I$ since the same number of tokens cannot be assigned to fewer experts.
\end{itemize}

It is worth noting that starting from $N_F \ge \text{TopK}$, $B_{\text{rank}}$ no longer increases, which implies that \textbf{the FLOPs of the FFN stage has reached its upper limit}.

This method provides a theoretical perspective for measuring $N_F$ to achieve extreme GEMM OFU. In practice, due to different padding and tiling behaviors in GEMM implementations, there is a certain FLOPS loss regardless of the arithmetic intensity. As shown in Figure~\ref{fig:roofline}, we conducted GEMM unit tests at different input scales in scenarios of balanced experts (i.e., experts having the same number of tokens) and imbalanced experts. The results show that although the overall HFU trend tends to flatten after the arithmetic intensity exceeds the platform's roofline ridge, a larger input scale can still improve the operator HFU. For comparison, data from \cite{deepseekopensourceweekday62025} estimates an average of 740 tokens per expert, indicating an HFU of approximately 60\% considering EP imbalance.

\subsection{AFD Dead Zone in Pursuit of HFU}

\begin{figure}[t!]
    \centering
    \begin{subfigure}{0.48\textwidth}
        \centering
        \includegraphics[width=\linewidth]{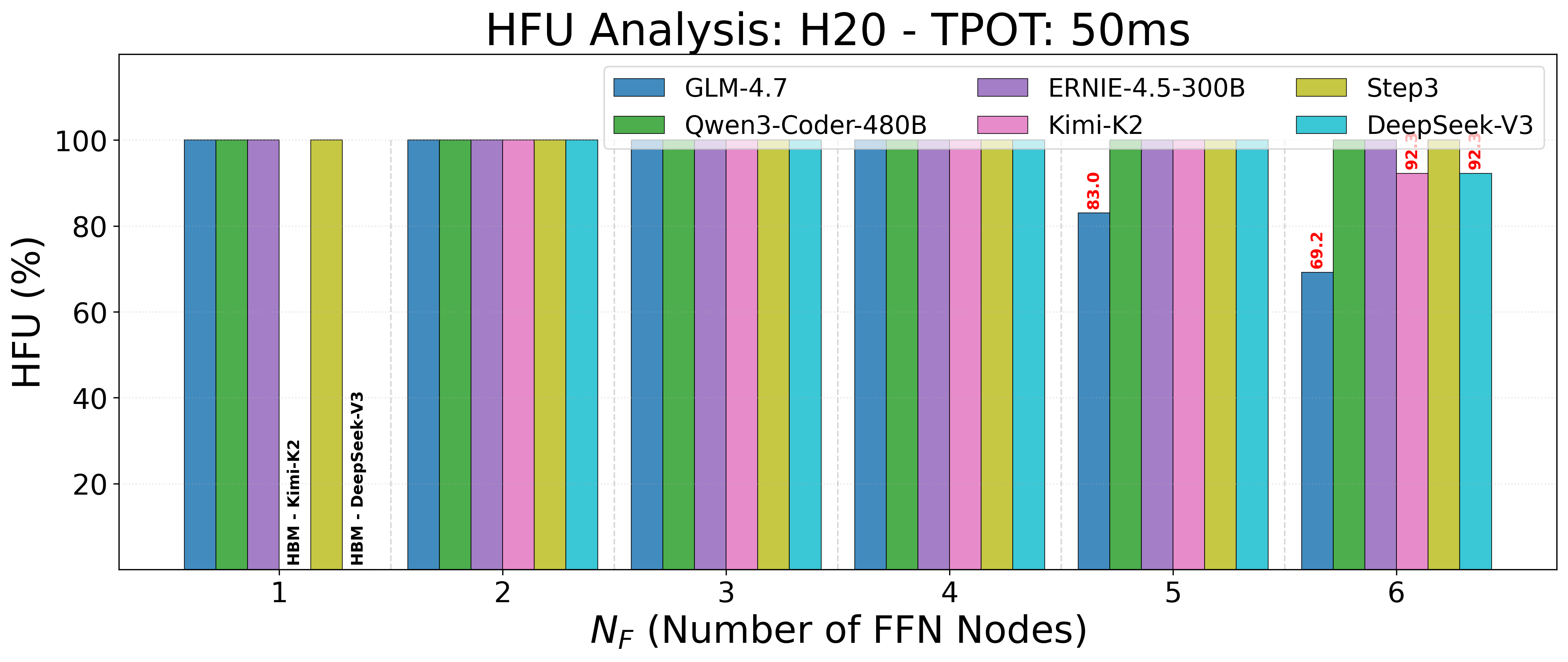}
        \caption{NVIDIA H20}
        \label{fig:HFU-1-H20}
    \end{subfigure}
    \hfill
    \begin{subfigure}{0.48\textwidth}
        \centering
        \includegraphics[width=\linewidth]{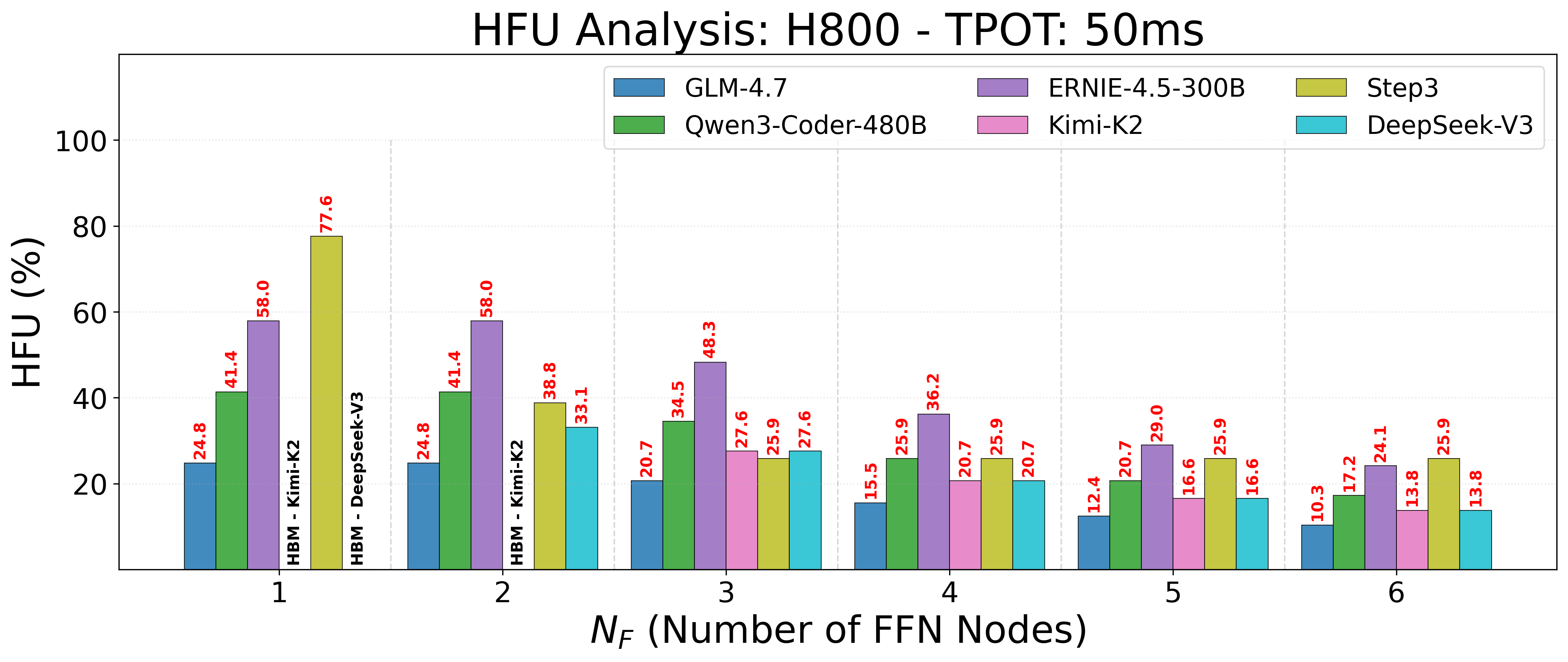}
        \caption{NVIDIA H800}
        \label{fig:HFU-2-H800}
    \end{subfigure}

    \vspace{2mm}

    \begin{subfigure}{0.48\textwidth}
        \centering
        \includegraphics[width=\linewidth]{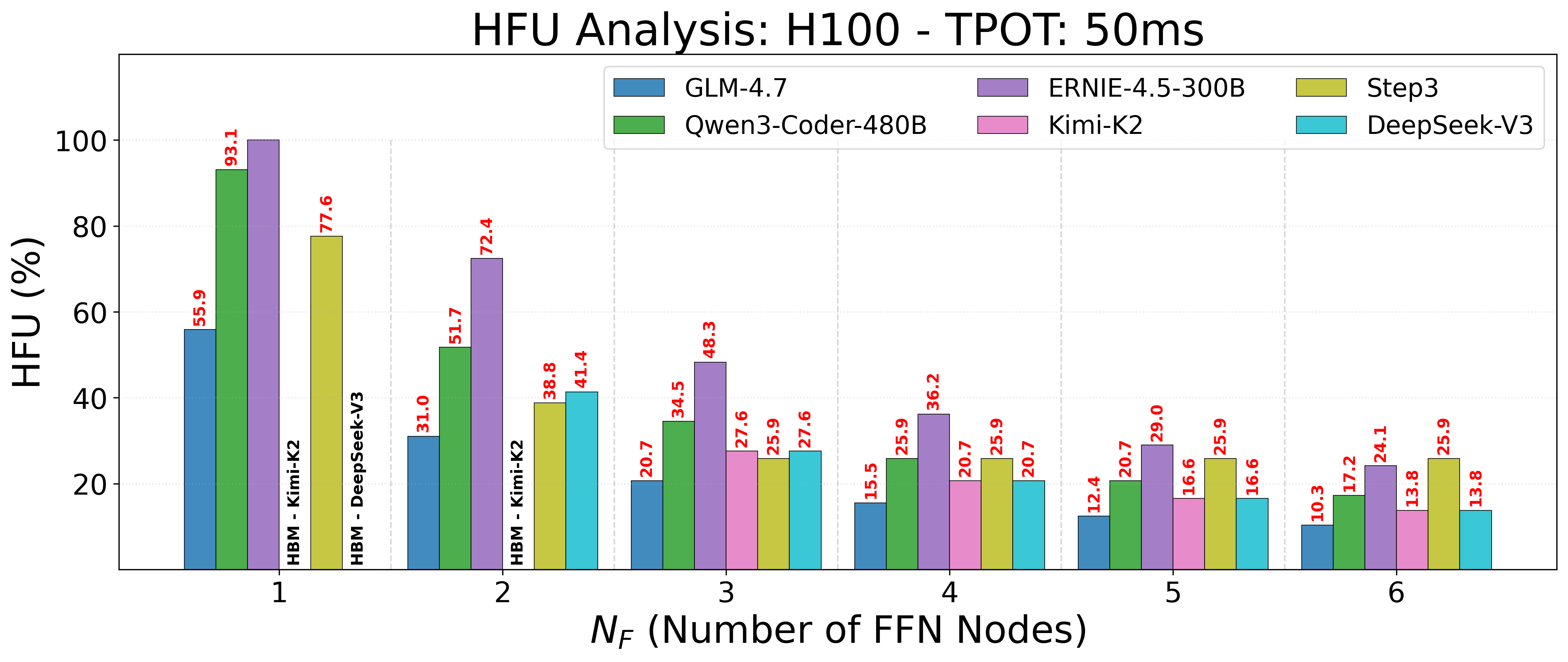}
        \caption{NVIDIA H100}
        \label{fig:HFU-5-H100}
    \end{subfigure}
    \hfill
    \begin{subfigure}{0.48\textwidth}
        \centering
        \includegraphics[width=\linewidth]{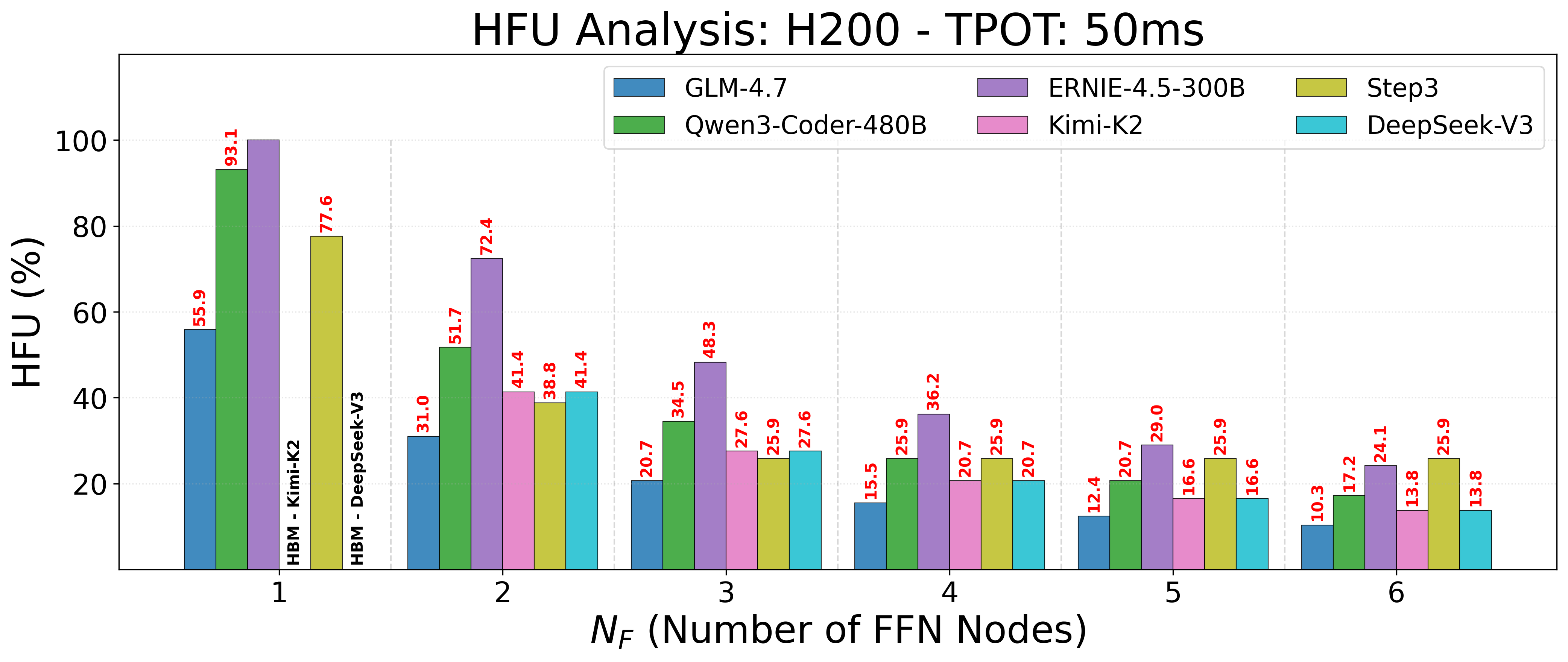}
        \caption{NVIDIA H200}
        \label{fig:HFU-6-H200}
    \end{subfigure}

    \vspace{2mm}

    \begin{subfigure}{0.48\textwidth}
        \centering
        \includegraphics[width=\linewidth]{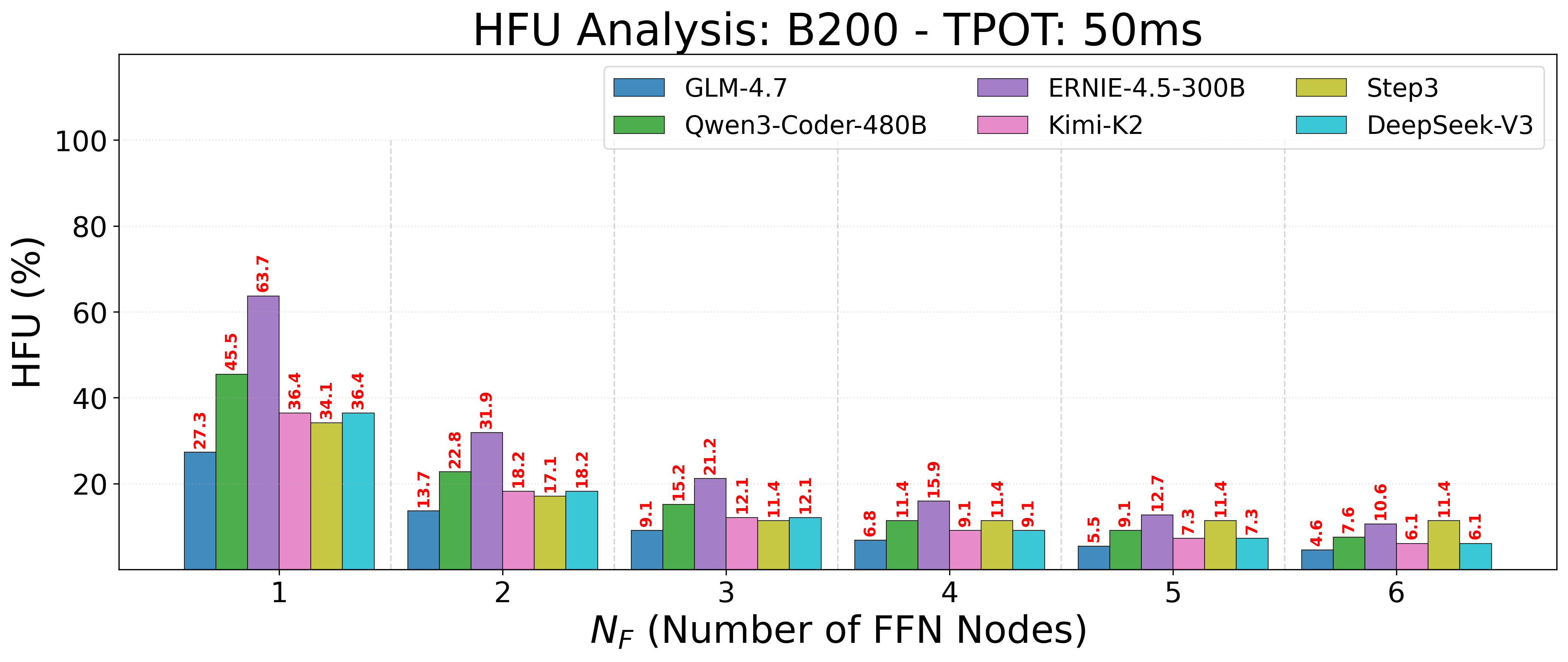}
        \caption{NVIDIA B200}
        \label{fig:HFU-3-B200}
    \end{subfigure}
    \hfill
    \begin{subfigure}{0.48\textwidth}
        \centering
        \includegraphics[width=\linewidth]{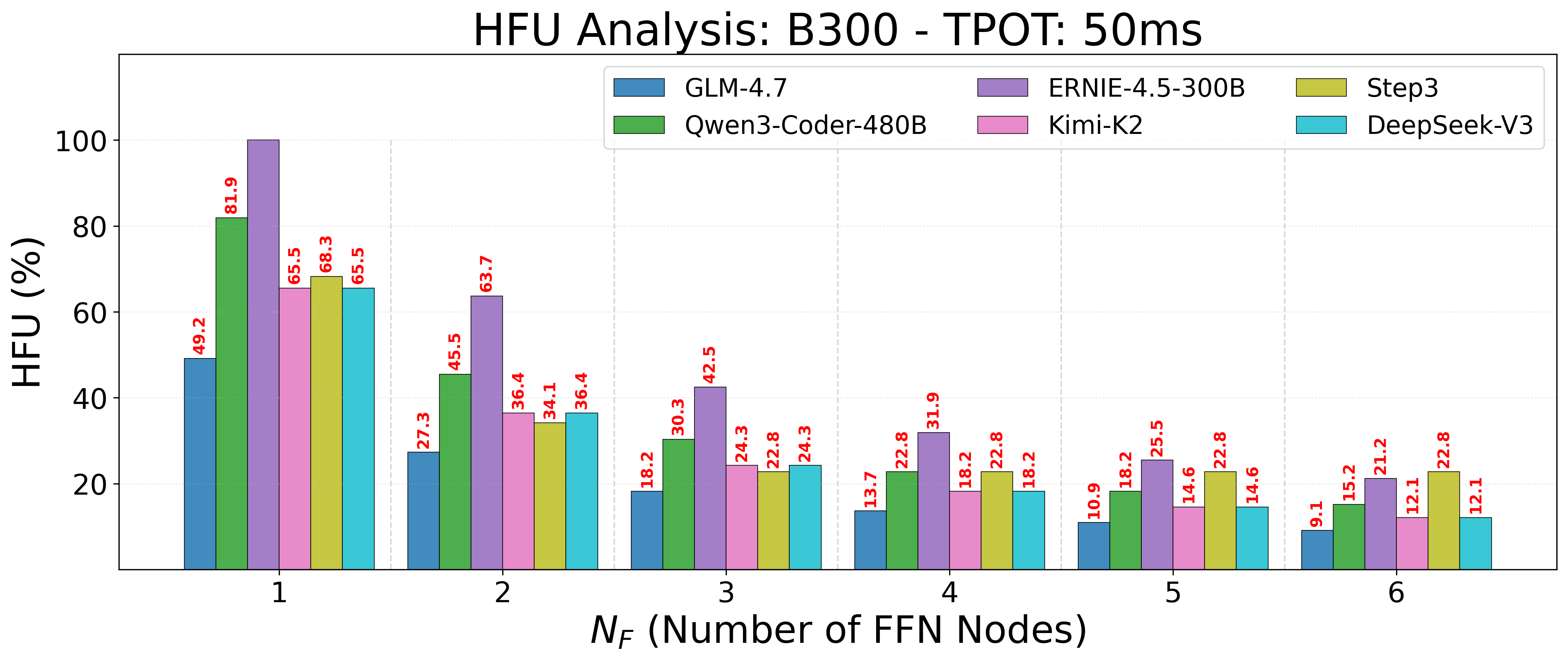}
        \caption{NVIDIA B300}
        \label{fig:HFU-4-B300}
    \end{subfigure}

    \vspace{2mm}

    \begin{subfigure}{0.48\textwidth}
        \centering
        \includegraphics[width=\linewidth]{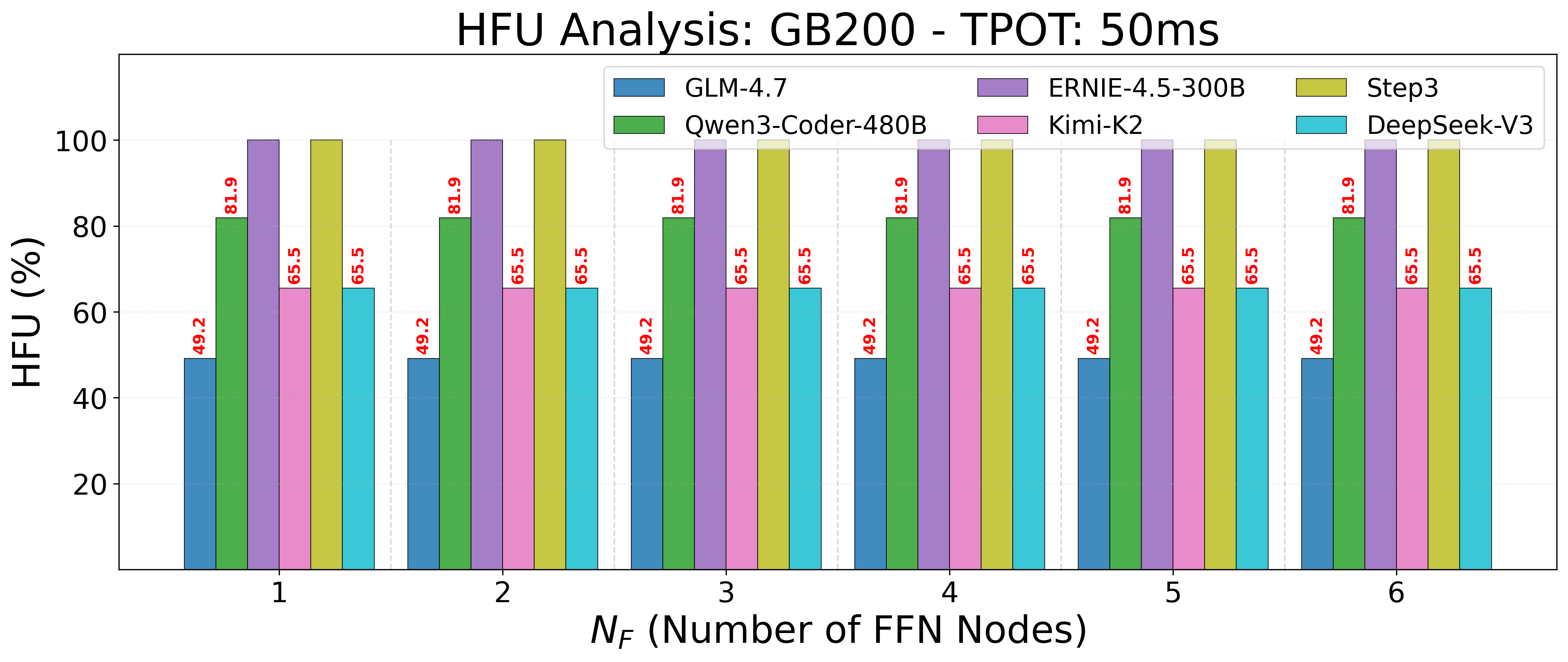}
        \caption{NVIDIA GB200}
        \label{fig:HFU-7-GB200}
    \end{subfigure}
    \hfill
    \begin{subfigure}{0.48\textwidth}
        \centering
        \includegraphics[width=\linewidth]{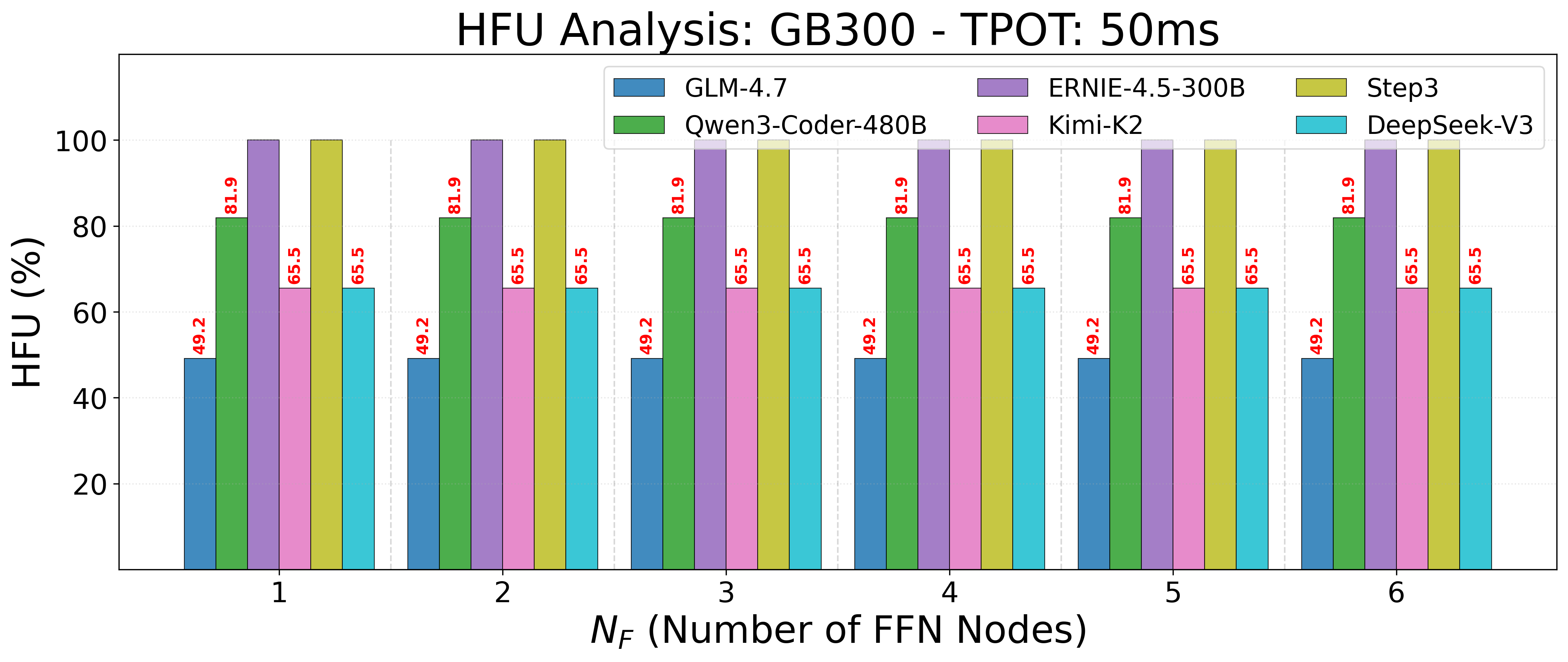}
        \caption{NVIDIA GB300}
        \label{fig:HFU-8-GB300}
    \end{subfigure}

   \caption{Theoretical upper-bound HFU of different models \cite{ernie2025technicalreport,kimiteam2025kimik2openagentic,qwen3technicalreport,5team2025glm45agenticreasoningcoding,deepseekai2025deepseekv3technicalreport,step3system} across hardware platforms under AFD deployment. The communication-bound HFU ceiling is highlighted in \textbf{\textcolor{red}{bold red}}. For all models, we assume MTP readiness with $L_{\text{accept}} = 1.7$, regardless of native support. All layers, dense or sparse, are assumed to have identical execution latency. We also assume a fixed deployment unit of 8 GPUs per node regardless of the physical Superpod scale. Detailed model and hardware configurations are provided in Appendix~\ref{appendix:model_and_hardware_configurations}.}
    \label{fig:hfu_for_models}
\end{figure}

By jointly applying the roofline model and Equation~\ref{eq:b_rank}, we plotted Figure~\ref{fig:hfu_for_models} to showcase the upper-bound HFU of different models on various hardware platforms. In the environment and scenario parameter assumptions, we set $L_{accept} = 1.7$ and assumed a gap of $t_g = 15$\,ms. The total FLOPs of the two grouped GEMM operators is calculated by $\text{FLOPs} = 6 \times G \times B \times H \times M$, while the memory access is $\text{Mem} = 3 \times G \times H \times M$. Here, $G$ represents the number of groups (i.e., the number of local experts), and $B = B_{\text{rank}} / G$ represents the average number of tokens obtained per expert.

Models with large parameter counts, such as DeepSeek-V3 and Kimi-K2, cannot be deployed on one or two 8-GPU nodes on NVIDIA Hopper series platforms, either limited by memory capacity or memory bandwidth. We identify them on the chart with text such as ``HBM - DeepSeek-V3''. Overall, Figure~\ref{fig:hfu_for_models} presents the following results:

\begin{itemize}
    \item \textbf{At the hardware level}, GPU platforms with relatively weak FLOPS (e.g., H20) or Superpod types equipped with abundant network bandwidth (e.g., GB200/GB300) can achieve higher theoretical HFU. For the former, high HFU is attainable with modest token counts due to the lower computational ceiling; for the latter, the abundant interconnect bandwidth can supply sufficient tokens to saturate the high FLOPS capacity. Conversely, on platforms with strong FLOPS but without a fully-interconnected scale-up network, models often exhibit worse theoretical HFU under AFD.
    \item \textbf{At the model level}, models with smaller sparsity and coarser expert granularity often yield higher theoretical HFU upper limits under AFD. For example, ERNIE-4.5 has a sparsity of 8 and an expert granularity of $8/3.5$, while for Step3, the sparsity is 16 and the expert granularity is $7/5$. These characteristics jointly allow them to generate more FLOPs requirements within limited inbound tokens.
    \item \textbf{As $N_F$ increases}, HFU does not continue to improve. Although larger $N_F$ raises arithmetic intensity (and thus OFU), the network bandwidth constraint causes $S_t$ of FFN to keep decreasing, keeping HFU at a low level.
\end{itemize}

To measure the relative benefit of AFD on standard clusters, it is also necessary to evaluate the theoretical difference compared to large-scale EP deployment. As analyzed in \S\ref{sec:arithmetic_intensity}, even if 100\% HFU cannot be achieved, large-scale EP deployment can still obtain an HFU of approximately 60\%. At this point, the theoretical HFU upper limit of AFD on non-Superpod H800 platform is only 33.1\%, where 100\% OFU is assumed. It is thus difficult for AFD to obtain a GEMM-stage HFU advantage compared to large-scale EP deployment on such configurations. We discuss the conditions under which AFD may exhibit greater potential in \S\ref{sec:afd-implications}.

\subsection{Imbalance Penalty}
\label{sec:imbalance_penalty}

Another aspect that has not been well-studied is the penalty caused by imbalance in AFD mode. The effectiveness of a deployment strategy can be measured by the average throughput per node, and this section uses this metric to compare AFD and large-scale EP parallelism from the perspective of DP imbalance and EP imbalance.

We define balancedness as $\sigma \le 1$, representing the correction factor for goodput loss in the attention or FFN stage influenced by imbalance. Let the batch sizes of the attention and FFN stages in the balanced state be $B_A$ and $B_F$, and the conversion factor (i.e., the penalty) for average throughput per node caused by imbalance be $\alpha \le 1$. Our task is to determine the relationship between $\alpha$ and $\sigma$ under the two deployment modes.

\subsubsection{DP Imbalance}

\begin{figure}[t!]
    \centering
    \begin{subfigure}[c]{0.48\textwidth}
        \centering
        \includegraphics[width=1.0\linewidth]{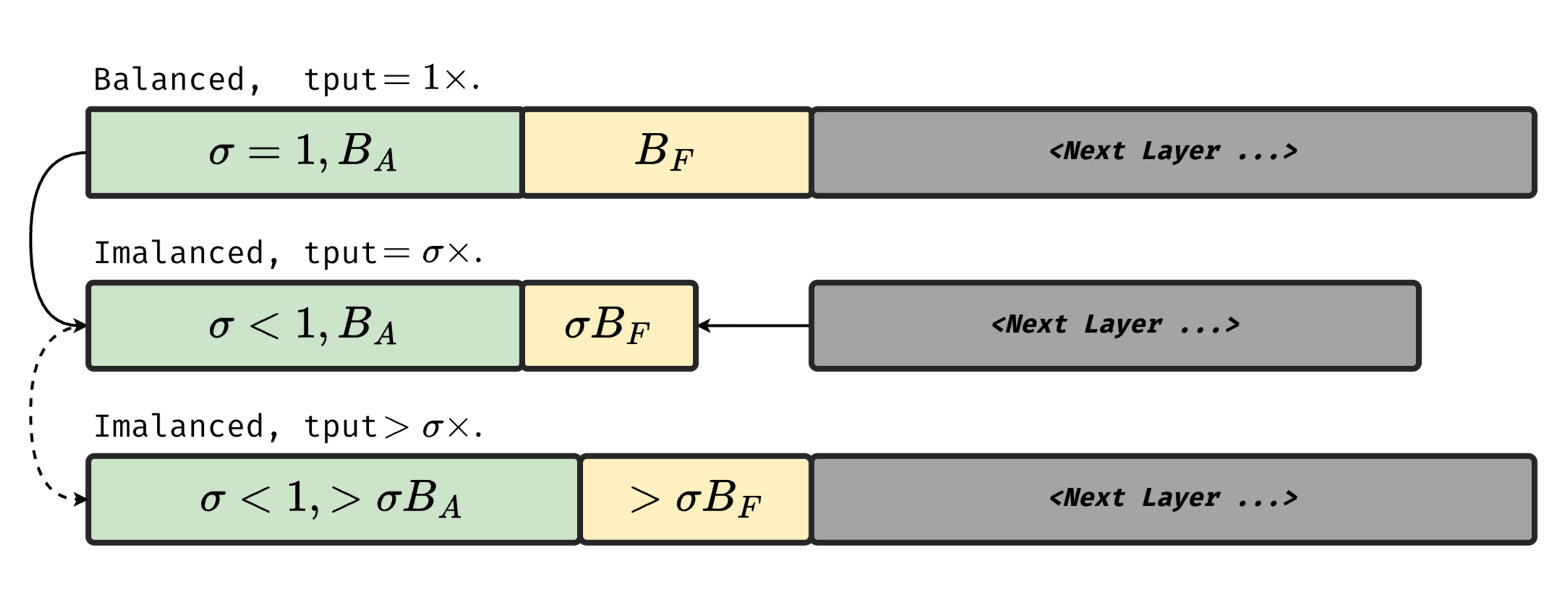}
        \caption{DP imbalance handling for large-scale EP. The decreased FFN latency from reduced batch size can be partially reclaimed by slightly increasing the batch.}
        \label{fig:dp_imbalance_ep}
    \end{subfigure}
    \hfill
    \begin{subfigure}[c]{0.48\textwidth}
        \centering
        \includegraphics[width=1.0\linewidth]{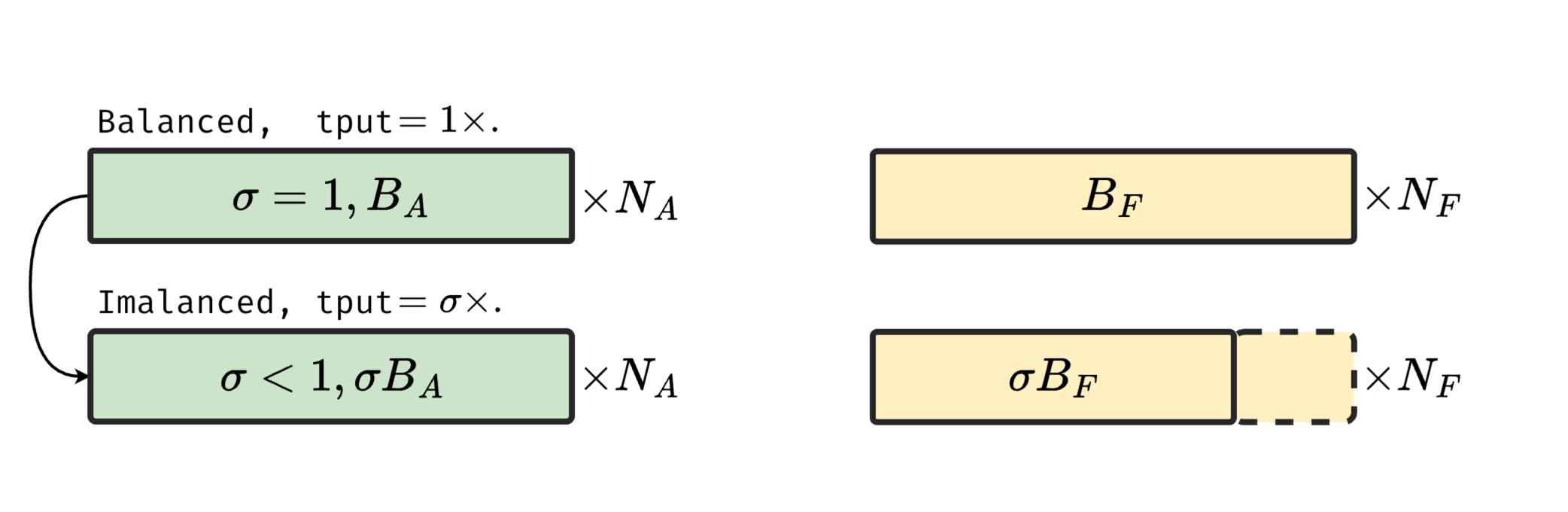}
        \caption{DP imbalance handling for AFD. The fixed $t_B$ budget prevents reclaiming latency released by the faster FFN stage.}
        \label{fig:dp_imbalance_afd}
    \end{subfigure}

    \vspace{2mm}

    \begin{subfigure}[c]{0.48\textwidth}
        \centering
        \includegraphics[width=1.0\linewidth]{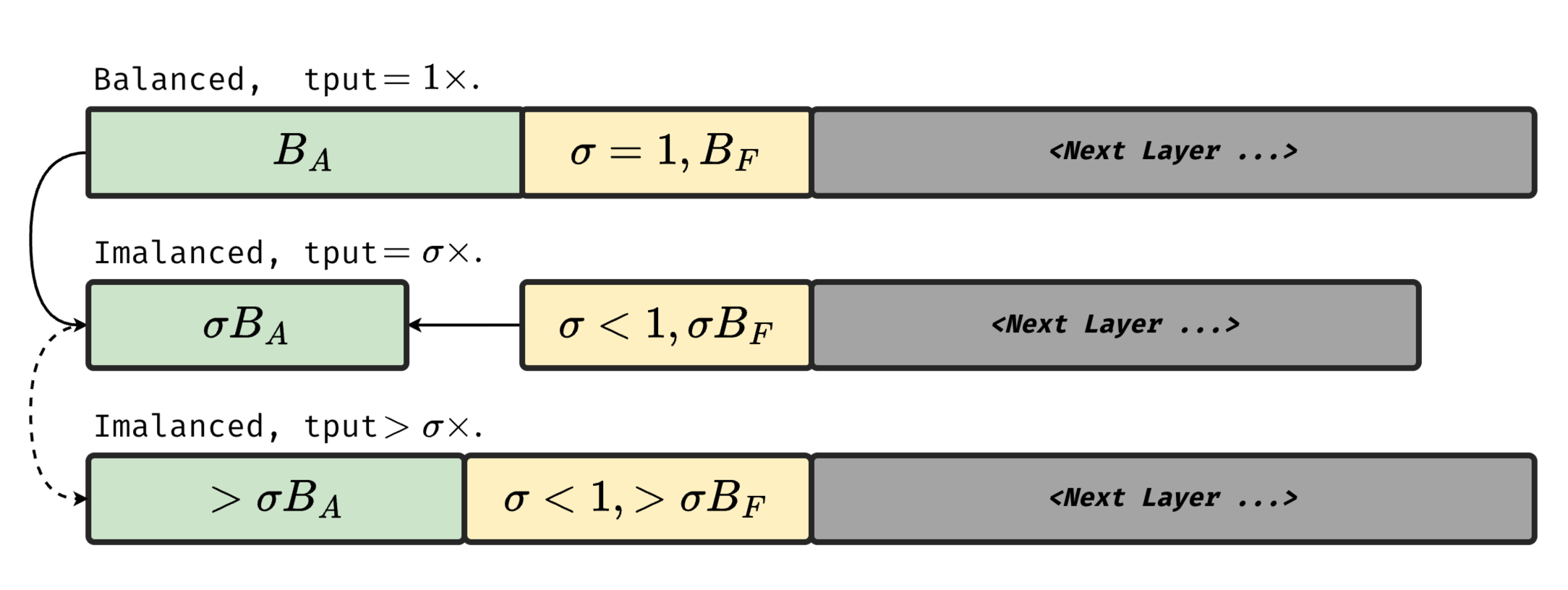}
        \caption{EP imbalance handling for large-scale EP. Continuous batch size adjustment allows fine-grained latency budget utilization.}
        \label{fig:ep_imbalance_ep}
    \end{subfigure}
    \hfill
    \begin{subfigure}[c]{0.48\textwidth}
        \centering
        \includegraphics[width=1.0\linewidth]{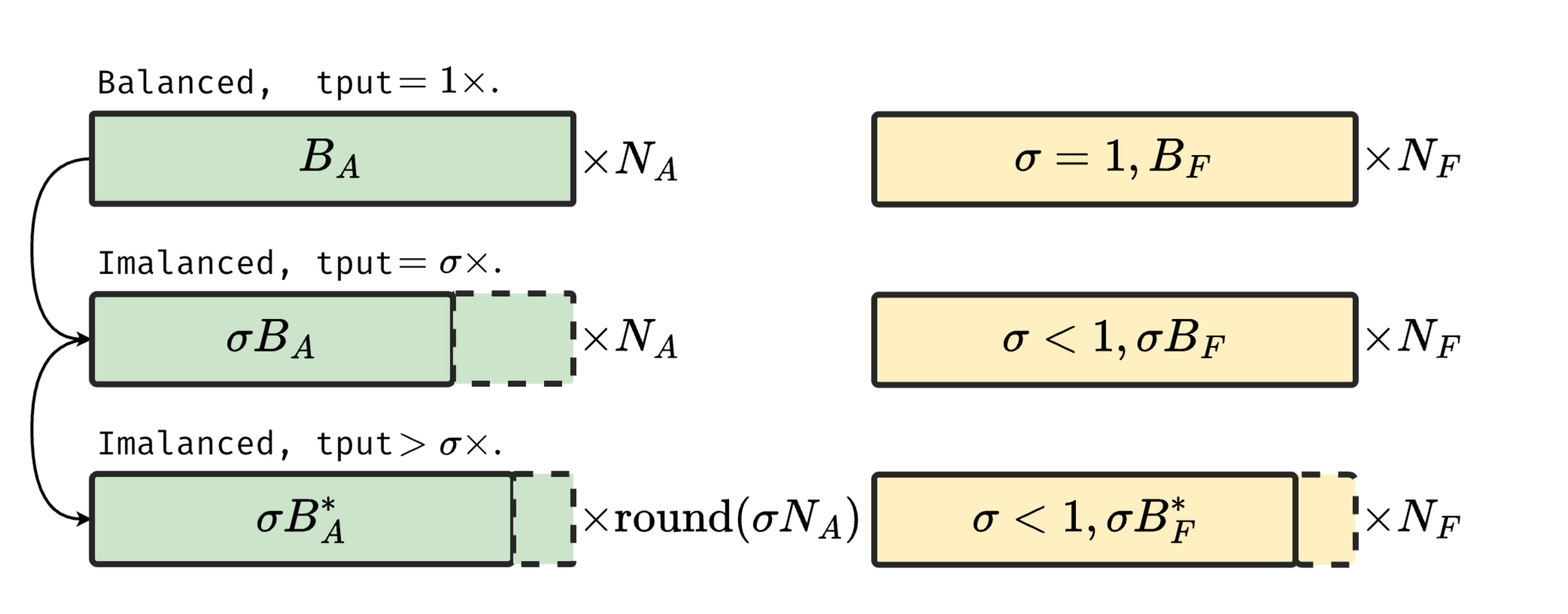}
        \caption{EP imbalance handling for AFD. Discrete $N_A$ scaling introduces quantization errors in resource allocation.}
        \label{fig:ep_imbalance_afd}
    \end{subfigure}

    \caption{Comparison of DP and EP imbalance handling strategies between large-scale EP and AFD deployments. The key distinction lies in AFD's discrete scaling constraints versus large-scale EP's continuous adjustment capability.}
    \label{fig:imbalance_sketch}
\end{figure}

DP imbalance stems from inconsistencies in context lengths and batch sizes across DP ranks, caused by varying input sequence lengths and the progression of requests during inference, resulting in inconsistent attention execution latencies. In real systems, the ideal attention execution latency under theoretical equilibrium will inevitably be exceeded when the total context length of requests surpasses a specific threshold. To prevent the SLO from being violated due to latency jitter caused by DP imbalance, the batch size must be reduced.

In large-scale EP deployment, when facing the DP imbalance problem, as shown in Figure~\ref{fig:dp_imbalance_ep}, the batch size is first reduced to $\sigma \times$ the original. At this point, according to our definition of the degree of imbalance, the latency of the attention part remains unchanged, while the batch of the FFN stage is reduced from $B_F$ to $\sigma B_F$, leading to a decrease in $t_f$. There are two further choices:

\begin{itemize}
    \item Maintain the status quo, i.e., introduce a throughput penalty of $\alpha_{\text{EP}} = \sigma$, but allow requests in existing instances to obtain smaller TPOT.
    \item Slightly increase the batch size to utilize the latency budget released due to the faster execution of the FFN stage. In this case, $\alpha_{\text{EP}} > \sigma$ is achieved.
\end{itemize}

In AFD, as shown in Figure~\ref{fig:dp_imbalance_afd}, the first step adopts the same strategy as in large-scale EP deployment, i.e., reducing the batch size to $\sigma$ times the original. However, due to the separation characteristics of AFD and the memory bottleneck on the FFN side, it is difficult to reuse the latency budget released by a smaller batch size. Consequently, $\alpha_{\text{AFD}} = \sigma$, indicating that the throughput penalty in AFD is greater than that of large-scale EP.

\subsubsection{EP Imbalance}
EP imbalance originates from the gating module in MoE selecting different subsets of routed experts for individual tokens, a phenomenon that cannot be eliminated under dynamic input. The direct consequence is a difference in communication and computation volume across EP ranks. For convenience of discussion, we ignore the case where communication becomes a bottleneck and only consider the impact of fluctuations on computation for the two deployment strategies and throughput metrics. The relationship of the latency metrics is formulated as:

\begin{itemize}
    \item For large-scale EP parallelism, $t_B = t_a + t_f$
    \item For AFD, $t_B = t_a = t_f$
\end{itemize}

\textbf{Large-scale EP deployment:} Assume that when the average context length is fixed, $t_a$ is proportional to batch size. When the batch is reduced to $\sigma \times$ in the first step, under the imbalance scenario, the total latency under the batch size $\sigma B_A$ becomes $\sigma t_a + t_f < t_B$. We assume that the batch size can be additionally increased to $b > \sigma B$, during which both $t_a$ and $t_f$ change linearly with the increase of $b$. Then we have:

\begin{align}
    \frac{b}{B}t_a + \frac{b}{\sigma B}t_f & = \alpha t_a + \frac{\alpha}{\sigma}t_f = t_B = t_a + t_f \\
    \alpha_\text{EP} & = \frac{t_a / t_f + 1}{t_a / t_f + 1 / \sigma} = \frac{\lambda_\text{EP} + 1}{\lambda_\text{EP} + 1 / \sigma} > \sigma
    \label{eq:alpha_ep}
\end{align}

The penalty of EP imbalance on large-scale EP deployment given by Equation~\ref{eq:alpha_ep} is a monotonically increasing function of the ratio $\lambda_\text{EP} = t_a / t_f$ in the latency budget. In practice, e.g., on the H800 platform, the value is typically $\lambda_{\text{EP}} \in [2, 4]$, and in the profile data provided by DeepSeek \cite{deepseekprofile2025}, $\lambda_{\text{EP}} \approx 4$. Furthermore, it should be noted that in the derivation, we assumed $\frac{b}{\sigma B} t_f$ as the corrected $t_f$. Due to the characteristic of increasing computational density, the accurate $t_f$ is a monotonically increasing convex function of batch size, implying that we actually \textbf{overestimated} the proportion of $t_f$. Therefore, the conversion factor $\alpha _\text{EP}$ caused by EP imbalance under large-scale EP deployment is larger than that in Equation~\ref{eq:alpha_ep}.

\textbf{AFD deployment:} To cope with EP imbalance, we need to adjust $N_A$ and the batch size of attention ranks to match the tokens across both ends. When $\sigma \times N_A$ is an integer, we can simply adjust $N_A$, and attention ranks remain in the optimal full-load state as originally planned. The penalty can hence be represented by the ratio of the average throughput per node under (1) imbalanced and (2) fully balanced conditions. The actual penalty coefficient becomes the ratio of the proportions of $N_A$ to $N_A + N_F$ before and after, i.e.:

\begin{equation}
    \alpha_\text{exact} = \frac{\frac{\sigma \times N_A}{\sigma \times N_A + N_F}}{\frac{N_A}{N_A + N_F}} = \sigma \frac{N_A + N_F}{\sigma N_A + N_F} = \frac{\lambda_\text{AFD} + 1}{\lambda_\text{AFD} + 1 / \sigma} > \sigma
    \label{eq:alpha_exact}
\end{equation}

Note that Equation~\ref{eq:alpha_exact} is almost identical to Equation~\ref{eq:alpha_ep}, except that the ratio changes from $\lambda_\text{EP} = t_a / t_f$ in large-scale EP deployment to $\lambda_\text{AFD} = N_A / N_F$ here. However, with the same formulaic expression, in large-scale EP, continuous fine-tuning on batch size can be adopted to control execution latency. In AFD, the means of adjusting $N_A$ must also be considered. When $\sigma \times N_A$ is not an integer, the optimal value after discrete scaling needs to be selected. The $\alpha_{\text{AFD}}$ can then be derived as:

\begin{align}
    \alpha_\text{floor} & = (\lambda_\text{AFD} + 1) \frac{\lfloor\sigma \times N_A\rfloor}{\lfloor\sigma \times N_A\rfloor + N_F} \\
    \alpha_\text{ceil} & = (\lambda_\text{AFD} + 1) \frac{\lceil\sigma \times N_A\rceil}{\lceil\sigma \times N_A\rceil + N_F} \times \frac{\sigma \times N_A}{\lceil\sigma \times N_A\rceil} \\
    \alpha_\text{AFD} & =
    \begin{cases}
    \alpha_\text{exact}, & \sigma \times N_A \in \mathbb{Z} \\
    \max(\alpha_\text{floor}, \alpha_\text{ceil}), & \sigma \times N_A \notin \mathbb{Z}
    \end{cases}
\end{align}

\begin{figure}[t!]
    \centering
    \includegraphics[width=1.0\linewidth]{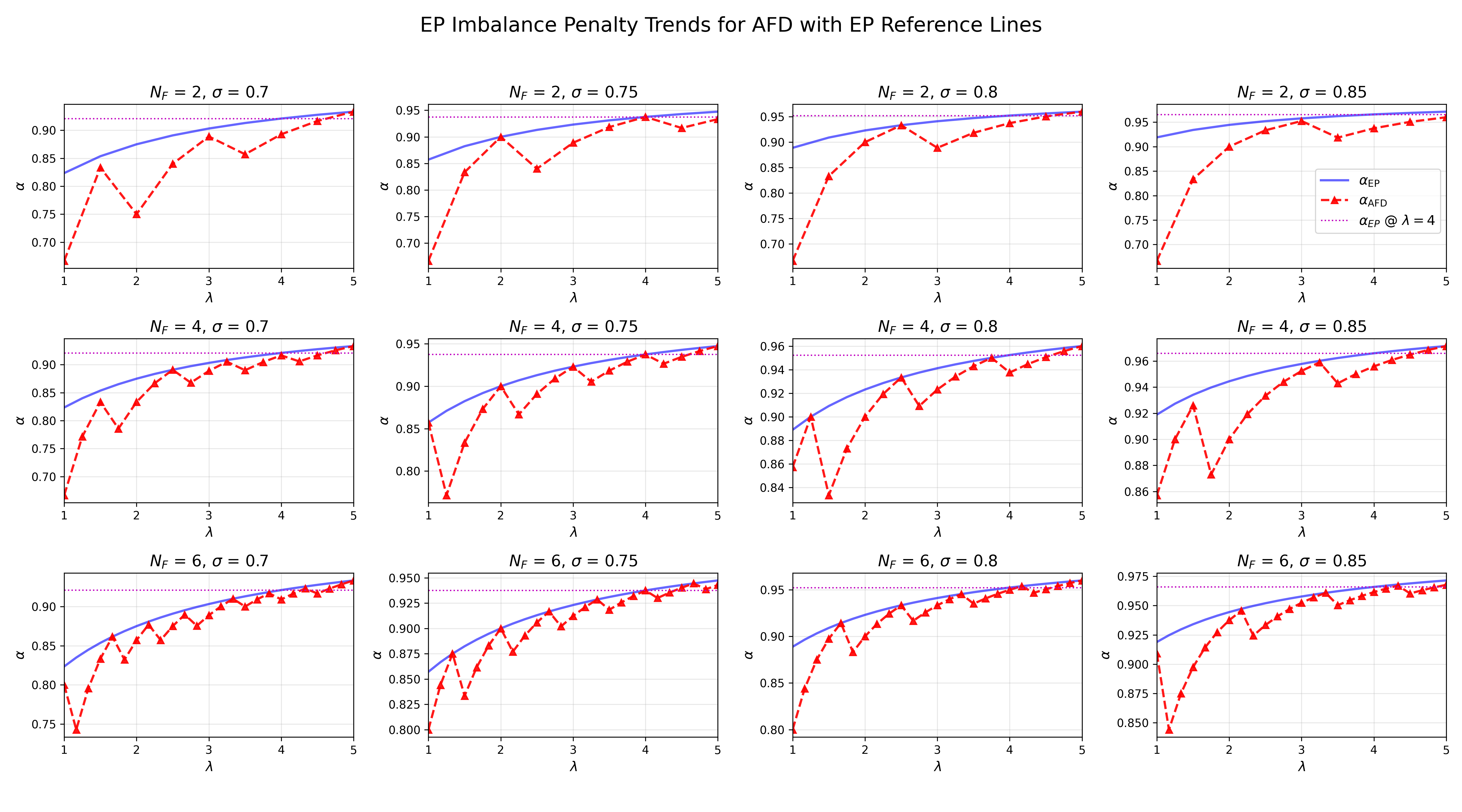}
    \caption{Throughput penalty from EP imbalance for AFD (red dashed curves) compared to large-scale EP (blue solid curves). We take $N_F \in \{2, 4, 6\}$, $\sigma \in \{0.7, 0.75, 0.8, 0.85\}$, and $\lambda \in [1, 5]$. The discrete scaling constraints in AFD result in worse penalties in most configurations.}
    \label{fig:penalty-trends}
\end{figure}

Note that in the calculation of $\alpha_\text{ceil}$, since attention batch size is limited by the finite FFN stage capacity when adopting the ceil-rounding method, an extra correction factor of $\frac{\sigma \times N_A}{\lceil\sigma \times N_A\rceil}$ is needed, as attention ranks are currently in underutilized status. We plotted the trends of $\alpha_\text{AFD}$ in Figure~\ref{fig:penalty-trends}. The figure indicates that, due to the problem of discrete scaling under AFD, it performs worse than large-scale EP in most cases.

It should be noted that $\lambda$ is actually an empirical measure of the ratio of computation latency between attention and FFN. Under large-scale EP deployment, it manifests as the ratio of duration occupied by both in the same operator stream. In AFD where attention and FFN are separately deployed and share the same $t_B$, the ratio becomes the ratio of $N_A$ to $N_F$. Figure~\ref{fig:penalty-trends} sets the $\lambda$ upper limit to 5, assuming that AFD can optimize computational efficiency for the FFN stage so that the FFN execution latency proportion is reduced from $1 / (1 + 4) = 0.2$ in the original large-scale EP deployment to $1 / (1 + 5) = 0.17$ in AFD. But even with such minor optimization objective, only when $\sigma$ exactly equals 0.8 can it barely achieve a consistent imbalance penalty with large-scale EP deployment. Therefore, the relationship between the expected profit margin and the penalty caused by discrete scaling should be carefully weighed when considering AFD.

\subsection{Implementation Challenges}
In addition to the performance risks theoretically analyzed earlier that AFD may cause due to bandwidth bottlenecks and discretization problems, various factors affecting the latency budget need to be considered in real systems. For example, technologies like CUDA Graph rely on PCIe communication, which may lead to prolonged graph launch latencies under a large operator count (e.g., 3BO will bring about $1.5\times$ the operator count on the attention side). Such a budget loss makes the assumption of $t_g = 15$\,ms somewhat optimistic. A smaller resulting $t_B$ directly constrains the overall batch size and may cause the FFN side to transition from \textit{compute-bound} to \textit{memory-bound}.

Furthermore, when discussing arithmetic intensity and HFU, we assumed that the value is linearly correlated to the token quantity per expert, and that the token throughput carried by the interconnect is also proportional to $t_B$, making the conclusion seemingly independent of its absolute value. However, in real systems, when the absolute $t_B$ is small, minor operators and boundary overheads account for a larger fraction of the overall cost, making them no longer negligible.

When AFD is deployed under a small number of $N_F$, the operator stream details of FFN also need to be further refined. For example, due to the different scales of attention and FFN instances and the communication pattern requirements for two-stage forwarding, additional shuffling operators are needed to conduct layout conversion from per source attention rank to per expert (and vice versa), and to allow the communication stream to run in parallel on an additional operator stream. These will further amortize the computational and bandwidth resources on the FFN side and lead to corrections on the FLOPS and memory bandwidth for bottleneck analysis. Currently, some research is attempting to solve similar problems on communication \cite{zhu2025fuscohighperformancedistributeddata}.

\section{Implications for AFD-Favorable Configurations}
\label{sec:afd-implications}

Based on the analysis presented in the preceding sections, we now synthesize the conditions under which AFD is more likely to yield performance benefits. Table~\ref{tab:afd_favorable_conditions} summarizes these conditions at the model and hardware levels.

\begin{table}[htbp]
\centering
\caption{Summary of conditions favorable for AFD deployment based on our analysis.}
\label{tab:afd_favorable_conditions}
\begin{tabular}{@{}l l p{6cm}@{}}
\toprule
\textbf{Dimension} & \textbf{Favorable Condition} & \textbf{Rationale} \\ \midrule
\multirow{2}{*}[0.5ex]{Model} & Coarse expert granularity & Larger $S_t$ and HFU potential \\ \cmidrule(l){2-3}
 & Lower expert sparsity & More tokens per EP rank \\ \midrule
\multirow{4}{*}[-1.5ex]{Hardware} & \multirow{3}{*}[-0.5ex]{Superpod-class architecture} & Sufficient bandwidth to saturate FFN \\
 & & Better baseline EP balance \\
 & & Reduced discrete scaling penalty \\ \cmidrule(l){2-3}
 & Heterogeneous resources & Targeted resource allocation \\ \bottomrule
\end{tabular}
\end{table}

Our analysis suggests that AFD exhibits greater potential in specific combinations of model configurations and hardware architectures, particularly in Superpod-class systems and models with coarse-grained experts.

\subsection{Model-Level Considerations}

At the model level, the following characteristics favor AFD deployment:

\begin{itemize}
    \item \textbf{Coarse-grained MoE experts}: Models with larger MoE intermediate sizes generate greater computational demand per token, improving temporal sparsity $S_t$ on the FFN side. As demonstrated in Appendix~\ref{appendix:moe_intermediate_size_to_hfu}, the theoretical HFU is directly proportional to the expert dimension $M$ when interconnect bandwidth becomes the bottleneck. For instance, Step3's expert dimension of 5120 yields substantially higher HFU potential compared to DeepSeek-V3 with $M=2048$.

    \item \textbf{Lower expert sparsity}: Expert sparsity, defined as the ratio $N_{experts}/\text{TopK}$, determines token distribution across EP ranks. Given a fixed expert size and the minimum latency for activating all local experts, the number of experts per rank is roughly determined. Consequently, the tokens that a single attention output can contribute to an EP rank is inversely proportional to expert sparsity. Models with lower sparsity exhibit workload distributions closer to dense models, improving per-expert batch sizes. For example, Step3 with a sparsity of $48/3 = 16$ achieves better token concentration than DeepSeek-V3 with a sparsity of $256/8 = 32$.
\end{itemize}

It is worth noting that these model-level characteristics run counter to the prevailing trend in MoE architecture design, which increasingly favors fine-grained experts and higher sparsity for improved training efficiency, as discussed in \S\ref{sec:trends_for_modern_moe}. This observation suggests that AFD may require co-design between model architecture and deployment infrastructure to realize its full potential.

\subsection{Hardware-Level Considerations}

At the hardware level, AFD benefits from systems possessing the following attributes:

\begin{itemize}
    \item \textbf{Superpod-class architecture}: Superpod systems such as NVIDIA GB200 and GB300 represent the most favorable hardware configuration for AFD. As shown in Figures~\ref{fig:HFU-7-GB200} and~\ref{fig:HFU-8-GB300}, such systems achieve substantially higher theoretical HFU under AFD. Superpod configurations offer three key advantages: (1) \textit{abundant interconnect bandwidth} from high-bandwidth, fully-interconnected scale-up networks that can supply sufficient tokens to saturate FFN FLOPS; (2) \textit{improved baseline EP balance} due to dense interconnect topology enabling more uniform token distribution across ranks; and (3) \textit{flexible resource partitioning} that breaks the node-granularity FFN scaling constraint of standard clusters, thereby reducing the discrete scaling penalty analyzed in \S\ref{sec:system-analysis}.

    \item \textbf{Heterogeneous resource availability}: When attention and FFN stages exhibit significantly different resource utilization patterns, AFD's physical separation enables targeted resource allocation. Operators can provision different hardware specifications for each role, potentially improving cost-efficiency compared to homogeneous deployments.
\end{itemize}

In conclusion, our analysis indicates that AFD holds greater promise in Superpod-class hardware systems and for models with coarse-grained expert configurations and lower expert sparsity. While current mainstream MoE architectures and standard cluster deployments may not fully benefit from AFD, the approach remains a valuable architectural option for next-generation infrastructure where these hardware constraints are alleviated.

\section{Discussion}

\subsection{TP Strategies}
Throughout the analysis of this paper, the parallelism mode on the attention side is limited to DP only, and that on the FFN side is limited to EP only. In fact, Tensor Parallelism (TP) is an option for both sides under AFD, as we discuss in this section.

\textbf{On the attention side}, the scale of attention (i.e., $N_A$) needs to be adjusted to accumulate the batch required by FFN. For scenarios serving longer contexts, a hybrid parallelism mode of DP + TP can be further employed on the attention side. However, this does not affect attention ranks' role of ``dynamic scaling, on-demand filling of FFN batch'' under AFD. For example, if TP-4 needs to be additionally introduced under the original DP scale to serve a $4\times$ context length scenario, maintaining the DP scale and expanding $N_A$ to $4\times$ for TP-4 will keep the FFN batch size unchanged. One should then carefully account for the synchronization overhead of request information and inference result reduction across TP ranks in the context of budget derivation.

\textbf{On the FFN side}, the ETP strategy (i.e., applying tensor parallelism at the expert stage) is effectively equivalent to adopting a finer-grained expert partitioning, as discussed in \S\ref{sec:trends_for_modern_moe}. This leads to reduced arithmetic intensity and therefore should be evaluated in conjunction with the $\text{TopK} / N_F$ ratio. When $\text{TopK} / N_F \ll B_{\text{ScaleUp}} / B_{\text{ScaleOut}}$, adopting ETP allows more effective utilization of the scale-up network bandwidth to compensate for the loss in arithmetic intensity. In this regime, a smaller EP degree improves load balance. However, it increases the number of partial results to be reduced, incurring additional overhead.

\subsection{AFD vs.\ PD Disaggregation}
Although both AFD and PD disaggregation are motivated by decoupling instances or stages with heterogeneous demands to pursue higher throughput, they differ fundamentally in nature.

In PD disaggregation, isolating the Prefill and Decode phases allows for independent optimizations on computation, memory footprint, and SLO requirements of each phase, without introducing strong coupling between the two types of instances. Modern infrastructures can effectively support elastic scaling of both phases and enable client-transparent request steering by dynamically establishing connections between them.

AFD, in contrast, can be viewed as a specialized pipeline-parallel (PP) deployment scheme. Its design introduces tighter stage synchronization between the two stages to promptly satisfy stringent TPOT requirements at the scale of tens of milliseconds. To the best of our knowledge, performance improvements in inference systems typically stem either from higher operator efficiency (e.g., operator fusion or increased arithmetic intensity) or from reducing GPU bubbles that are otherwise difficult to exploit in current systems (e.g., improving overlap by lowering communication overhead). However, as revealed by our analysis, these objectives are not well achieved in AFD on standard cluster configurations:

\begin{itemize}
    \item Compared to standard large-scale EP deployments, the FFN stage in AFD struggles to attain genuinely higher HFU due to limited interconnect bandwidth on non-Superpod platforms.
    \item Rather than directly optimizing explicitly existing GPU bubbles, AFD introduces a fragile 3BO pattern with a tight latency budget, along with the potential risk that latency jitter from DP imbalance and EP imbalance rapidly propagates across the attention and FFN stages.
\end{itemize}

\subsection{Expert-as-a-Service Architectures}
On the FFN side, all inputs and outputs are presented in the form of tokens and metadata, and directly interfaced with the AFD communication library, making it essentially a stateless component in LLM inference. The simple input, compute, and output stream of FFN has spawned deployment modes that further decouple FFN---especially MoE layers---from attention computation in a service-oriented form, namely Experts-as-a-Service (EaaS) \cite{liu2025expert,xiao2025xdeepserve}. Such an architecture, with fully decoupled A-F communication, is expected to allow one expert instance to serve a larger attention instance.

The practical viability of EaaS depends on several factors that require careful consideration. Due to strict SLO limits for online services, methods similar to micro-queuing \cite{wang2025toward} that prioritize serving local experts reaching buffer thresholds may not guarantee that forward propagation of requests in the current MoE layer will be completed within the expected latency. This method essentially trades latency metrics for improved MoE stage throughput metrics.

The scale requirements for EaaS deployments vary significantly with model size. For large-scale models such as DeepSeek-V3, activating all experts simultaneously would require substantial infrastructure (approximately $58\times$ the base node count) due to the asynchronous nature of DP in EaaS, where inbound FFN tokens may expect to activate experts on arbitrary layers. In such configurations, accumulating sufficient tokens for effective MoE forward pass necessitates a correspondingly large attention-side deployment.

In contrast, EaaS architectures are considerably more viable for smaller models, which exhibit manageable memory access latencies and more modest FLOPS requirements. The reduced infrastructure requirements render the inherent trade-offs more favorable, suggesting that service-oriented expert deployment may initially gain traction in smaller-scale scenarios before extending to larger deployments as infrastructure matures.

\section{Conclusion}
In this paper, we conducted a systematic evaluation of Attention-FFN Disaggregation (AFD) for modern MoE deployment. By establishing a budget-based analytical framework, we revealed that AFD's performance characteristics vary significantly across different hardware configurations and model architectures.

On standard cluster configurations, our analysis highlights a critical ``dead zone'' where scale-out bandwidth bottlenecks prevent FFN instances from achieving higher HFU despite scaling. We further demonstrated that AFD's discrete node-level scaling makes it inherently less robust to DP and EP load imbalances compared to the continuous batch adjustment capabilities of large-scale EP.

However, our analysis also identifies specific conditions under which AFD exhibits greater potential. Superpod-class hardware systems can effectively alleviate these constraints through abundant interconnect bandwidth, improved EP balance from dense interconnect topology, and flexible resource partitioning that reduces discrete scaling penalties. On the model side, coarse-grained expert configurations and lower sparsity generate higher computational demand per token, improving HFU potential under AFD deployment.

While the industry trend toward fine-grained experts presents challenges for AFD on current mainstream hardware, our findings position AFD as a promising approach for specific hardware-model combinations rather than a universal solution. We hope this work provides a rigorous theoretical foundation for architectural decisions in large-scale model serving.

\printbibliography

\clearpage
\appendix
\section{MoE Intermediate Size vs.\ HFU}
\label{appendix:moe_intermediate_size_to_hfu}

Readers may have observed that in Figures~\ref{fig:HFU-7-GB200} and~\ref{fig:HFU-8-GB300}, models Kimi-K2 and DeepSeek-V3 exhibit an identical theoretical HFU upper bound. This section provides an analytical derivation for this phenomenon.

We denote the scale-up network bandwidth of the Superpod as $\mathcal{B}_{\text{ScaleUp}}$ and the model's expert dimension (i.e., MoE intermediate size) as $M$. In a Superpod environment, the following relationship holds:

\begin{equation}
B_{\text{rank}} = B_{\text{ScaleOut}} = B_{\text{ScaleUp}} = \frac{\mathcal{B}_{\text{ScaleUp}} \times t_B}{(\text{sizeof}(\text{fp8}) + \text{sizeof}(\text{bf16})) \times H} = \frac{\mathcal{B}_{\text{ScaleUp}} \times t_B}{3H}
\end{equation}

For a given number of input tokens, the computational workload of the grouped GEMM operations is:

\begin{equation}
\text{FLOPs} = 6 \times G \times B \times H \times M = 6 \times B_{\text{rank}} \times H \times M = 2 \times \mathcal{B}_{\text{ScaleUp}} \times M \times t_B
\end{equation}

Accordingly, the HFU within the $t_B$ window can be derived as:

\begin{equation}
\text{HFU} = \frac{\text{FLOPs}}{\text{FLOPS} \times t_B} = \frac{2 \times \mathcal{B}_{\text{ScaleUp}} \times M}{\text{FLOPS}}
\label{eq:hfu_to_m}
\end{equation}

Since these models are primarily constrained by the interconnect bottleneck in this specific case, the theoretical achievable HFU depends exclusively on the model parameter $M$. For NVIDIA GB200/GB300 systems, where $\text{FLOPS} = 4500$\,TFLOPS and $\mathcal{B}_{\text{ScaleUp}} = 720$\,GB/s, Equation~\ref{eq:hfu_to_m} yields exactly 65.5\% theoretical HFU for these models. Similarly, all other enumerated models reach the maximum theoretical HFU, with the exception of GLM-4.7, which exhibits a lower value due to its smaller $M=1536$.

\section{Model and Hardware System Configurations}
\label{appendix:model_and_hardware_configurations}

In this section, we provide the related configurations for the models and hardware systems used in this study, respectively shown in Tables~\ref{tab:model_configurations} and~\ref{tab:hardware_configurations}.

\begin{table}[htbp]
\centering
\caption{Model Configurations}
\label{tab:model_configurations}
\resizebox{\textwidth}{!}{
\begin{tabular}{lcccccc}
\toprule
\textbf{Configuration Parameter} & \textbf{DeepSeek-V3} & \textbf{Kimi-K2} & \textbf{Step3} & \textbf{Qwen3-Coder} & \textbf{ERNIE-4.5} & \textbf{GLM-4.7} \\
\midrule
Hidden Size             & 7168 & 7168 & 7168 & 6144 & 8192 & 5120 \\
Number of Layers        & 61 & 61 & 61 & 62 & 54 & 92 \\
Number of Dense Layers  & 3 & 1 & 5 & 0 & 3 & 3 \\
Number of MoE Layers    & 58 & 60 & 56 & 62 & 51 & 92 \\
Number of Routed Experts & 256 & 384 & 48 & 160 & 64 & 160 \\
Number of Experts per Token & 8 & 8 & 3 & 8 & 8 & 8 \\
MoE Intermediate Size   & 2048 & 2048 & 5120 & 2560 & 3584 & 1536 \\
\bottomrule
\end{tabular}
}
\end{table}

\begin{table}[htbp]
\centering
\caption{Hardware System Configurations}
\label{tab:hardware_configurations}
\resizebox{\textwidth}{!}{
\begin{tabular}{lcccccccc}
\toprule
\textbf{System Parameter} & \textbf{H20} & \textbf{H100} & \textbf{H200} & \textbf{H800} & \textbf{B200} & \textbf{B300} & \textbf{GB200} & \textbf{GB300} \\
\midrule
Peak FP8 Performance (TFLOPS)  & 296 & 1979 & 1979 & 1979 & 4500 & 4500 & 4500 & 4500 \\
Memory Bandwidth (TB/s)        & 4 & 3.35 & 4 & 3.35 & 7.7 & 8 & 7.7 & 8 \\
Memory Capacity (GB)           & 96 & 80 & 141 & 80 & 180 & 270 & 180 & 270 \\
Scale-Out Bandwidth (GB/s)     & 50 & 50 & 50 & 50 & 50 & 100 & Superpod & Superpod \\
Scale-Up Bandwidth (GB/s)      & 360 & 360 & 360 & 160 & 720 & 720 & 720 & 720 \\
\bottomrule
\end{tabular}
}
\end{table}

\end{document}